\documentclass[fleqn,usenatbib]{mnras}
\usepackage{newtxtext,newtxmath}
\usepackage[T1]{fontenc}
\usepackage{mathrsfs}
\usepackage{amsmath}
\usepackage{color}
\usepackage{mwe}
\usepackage{newtxtext,newtxmath}
\usepackage{ae,aecompl}
\usepackage{float}
\usepackage{graphicx}	
\usepackage{amssymb}	
\usepackage{xcolor}

\title[Variability in disk]{Relativistic, Viscous, Radiation Hydrodynamic Simulations of Geometrically Thin Disks. II. Disk Variability}
\author[Mishra et al.]{Bhupendra Mishra $^{1}$, Wlodek Klu\'zniak $^{2}$, P. Chris Fragile $^{3}$ \\
$^{1}$ JILA, University of Colorado and National Institute of Standards and Technology, 440 UCB, Boulder, CO 80309-0440, USA \\
$^{2}$ Nicolaus Copernicus Astronomical Center, Polish Academy of Sciences, Bartycka 18, Warsaw, 00-716, Poland \\
$^{3}$ Department of Physics and Astronomy, College of Charleston, Charleston, SC 29424, USA
}
\date{Accepted XXX. Received YYY; in original form ZZZ}

\pubyear{2018}

\begin{document}

\label{firstpage}
\pagerange{\pageref{firstpage}--\pageref{lastpage}}
\maketitle
\begin{abstract}
We perform detailed variability analysis of two-dimensional viscous, radiation hydrodynamic numerical simulations of Shakura-Sunyaev thin disks around a stellar mass black hole. Our accretion disk models are initialized on both the gas-, as well as radiation-, pressure-dominated branches of the thermal equilibrium curve, with mass accretion rates spanning the range from $\dot{M} = 0.01 L_\mathrm{Edd}/c^2$ to $10 L_\mathrm{Edd}/c^2$. An analysis of high-frequency temporal variations of the numerically simulated disk reveals multiple robust, coherent oscillations. Considering the local mass flux variability, we find an oscillation occurring at the maximum value of the radial epicyclic frequency, $3.5\times 10^{-3}\,t_\mathrm{g}^{-1}$, a possible signature of a trapped fundamental ${\it g}$-mode. Although present in each of our simulated models, the trapped ${\it g}$-mode feature is most prominent in the gas-pressure-dominated case. Additionally, the total pressure fluctuations in the disk suggest strong evidence for standing-wave ${\it p}$-modes with frequencies $\le 3.5\times 10^{-3}\,t_\mathrm{g}^{-1}$, some trapped in the inner disk close to the ISCO, others present in the middle/outer parts of the disk. Knowing that the trapped ${\it g}$-mode frequency and maximum radial epicyclic frequency differ by only $0.01\%$  in the case of  a non-rotating black hole, we simulated an additional initially gas-pressure-dominated disk with a dimensionless black hole spin parameter $a_* = 0.5$. The oscillation frequency in the spinning black hole case confirms that this oscillation is indeed a trapped ${\it g}$-mode. All the numerical models we report here also show a set of high frequency oscillations at the vertical epicyclic and breathing mode frequencies. The vertical oscillations show a 3:2 frequency ratio with oscillations occurring approximately at the radial epicyclic frequency, which could be of astrophysical importance in observed twin peak, high-frequency quasi-periodic oscillations. The comparison of inferred power among all the individual oscillations in our models suggests that the strongest oscillations occur at the breathing oscillation frequency.
\end{abstract}

\footnotetext[1]{Kavli Institute for Theoretical Physics, Santa Barbara, CA.}

\section{Introduction}

It has long been established that thin accretion disks are prone to oscillations, with several relativistic eigenmodes having been predicted by two groups  \citep{Kato80,Okazaki87, Nowak91, Nowak92, Nowak93} , prior to the first observations of high-frequency quasi-periodic oscillations (HFQPOs) in the X-ray lightcurves of black hole candidates \citep{Morgan97, Remillard99}.  HFQPOs at characteristic frequencies have since been detected from several luminous accreting neutron stars, as well as a few black hole candidates in X-ray binaries \citep[see][for reviews]{vanderKlis04,Remillard06}.

In this work, we distinguish between mode angular frequencies, $\omega$, along with
their corresponding radial epicyclic and orbital angular frequencies, $\kappa$ and $\Omega$, and the usual frequencies (reported by
observers in Hz), $\nu \equiv \omega/(2\pi)$, with the corresponding radial epicyclic and orbital frequencies, $\tilde{\kappa}\equiv \kappa/(2\pi)$ and $\tilde{\Omega} \equiv \Omega/(2\pi)$.
In the Schwarzschild metric, the maximum value of the radial epicyclic frequency,  
$\tilde{\kappa}_\mathrm{max} = 3.51686 \times 10^{-3}\,c^3/GM$, 
is attained at $r = r_\mathrm{max} = 8GM/c^2\equiv8\,r_\mathrm{g}$, 
and happens to satisfy $\tilde{\kappa}_\mathrm{max} = \tilde{\Omega}(r_\mathrm{max})/2$. In spherically symmetric gravity, so in the Schwarzschild metric---and the Paczy\'nski-Wiita pseudo-potential---the vertical epicyclic frequency, $\nu_\perp$, coincides with the orbital frequency $\tilde\Omega$. However, in general they are not equal, e.g., for prograde orbits in the Kerr metric, $\nu_\perp < \tilde\Omega$. 

{\bf \it Pressure waves or ${\it p}$-modes:} The fundamental (i.e., ones with no vertical nodes) ${\it p}$-modes (oscillation modes driven by pressure waves) 
are evanescent within the epicyclic frequency curve, i.e., whenever their observed angular frequency satisfies $\nu<\tilde{\kappa}(r)$ \citep{Kato2001}, where $\nu$ and $\tilde{\kappa}(r)$ are the oscillation and radial epicyclic frequencies. Therefore, they may either travel in the outer part of the disk [note that at large radii, $\kappa(r)$ tends to the Keplerian frequency $\Omega(r)$], or they may be trapped between the inner disk radius, say at the innermost stable circular orbit (ISCO), and that radius $r$, for which  $\nu=\tilde{\kappa}(r)$ \citep{Ortega02}.  As we will see in the rest of this article, such trapped ${\it p}$-modes are present in all of our simulations. We also observe ${\it p}$-modes for $r>8\,r_\mathrm{g}$.  

{\bf \it Gravity waves or ${\it g}$-modes: }
In contrast, the fundamental (gravity) ${\it g}$-modes are evanescent for $\nu>\tilde{\kappa}(r)$; hence, they remain trapped close to $r=r_\mathrm{max} = 8\,r_\mathrm{g}$ (for a non-rotating black hole), and their eigenfrequencies belong to a discrete spectrum. In the Schwarzschild metric, the eigenfrequency of the lowest-frequency ${\it g}$-mode is
$\sigma/(2\pi)= 3.35723\times 10^{-3}\,c^3/(GM)$, with the fundamental (no node, even parity) ${\it g}$-mode at the slightly higher frequency of
$\sigma_0/(2\pi)= 3.49552\times 10^{-3}\,c^3/(GM)$,
for a $0.1\,L_\mathrm{Edd}$ radiation pressure-dominated disk, and slightly higher at lower luminosities \citep{Perez97}. Thus, there should be no ${\it g}$-modes (from linear theory) with frequency $\nu < 3.357\times 10^{-3}\,c^3/(GM)$. 

{\bf \it Corrugation waves or ${\it c}$-modes: }
As our focus in this paper is primarily on non-rotating black holes, we do not discuss  ${\it c}$-modes, which require a splitting between the vertical epicyclic and orbital frequencies \citep{Silbergleit01}. 

While it seems clear that some HFQPOs must be a manifestation of eigenmodes of the accretion disk in general relativity (GR), their exact theoretical interpretation is as yet unclear. Initially, the linear theory of thin-disk oscillations in general relativity \cite[reviewed in][]{Wagoner99, Kato2001, Kato08} seemed quite promising \citep{Wagoner01}. However, after the discovery that the $\sim6\,M_\odot$ black hole candidate GRO J1655-40 exhibits a pair of HFQPOs at $300\,\mathrm{Hz}$ and $450\,\mathrm{Hz}$ \citep{Strohmayer01}, it has been noted that the two frequencies are in a 3:2 ratio \citep{Abramowicz01}, and a non-linear resonance model was developed to account for the ratio \citep{Kluzniak01, Kluzniak02, Kluzniak04, Bursa04, Kluzniak05}. Currently about four black holes are known to exhibit pairs of HFQPOs, in all cases in a 3:2 or 5:3 ratio, apparently \citep{Remillard02,Kluzniak02,Remillard04,Homan05}. 

It is natural to search for such periodicities in the temporal behavior of simulated global accretion disk models. High-frequency periodicities compatible with linear modes of diskoseismology have been found in several prior simulations. Early 1D (height-integrated) disk simulations revealed global oscillations (i.e., modes) at frequencies close to $\kappa_\mathrm{max}$, the maximum value of the radial epicyclic frequency $\kappa$ \citep{Honma92,Chen95, Milsom96}. Similar oscillations have been found in a two-dimensional simulation \citep{Milsom97}. In all cases these oscillations have been associated with (pressure-acoustic) ${\it p}$-modes.

\citet{ONeill09} have studied the time variability in a 2.5-D Newtonian simulation of a geometrically and optically thin polytropic $\alpha$-disk using the Paczynski-Wiita pseudo-potential, which mocks up qualitative effects of the Schwarzschild metric, such as the marginally stable orbit and a local maximum in $\kappa(r)$. Most of their simulations were initialized with a vertically perturbed disk, specifically one with its initial height exceeding the equilibrium height. In the sense that our radiation-pressure dominated initial setup runs collapse into gas-pressure-dominated disks in equilibrium \citep[see Fig. 1 in ][]{Fragile18}, they resemble the \citet{ONeill09} simulations, so perhaps it is not surprising that we recover similar results, albeit at a lower value of the viscosity parameter, $\alpha=0.02$, than their comparable $\alpha=0.1$ results.
Generally, the power density spectra of the vertically perturbed disks of \citet{ONeill09} are somewhat noisy, but the authors were able to infer global oscillations at a frequency close to $\kappa_\mathrm{max}$ of the Paczynski-Wiita pseudo-potential, which they interpreted as trapped ${\it g}$- and/or ${\it p}$-modes, as well as traveling ${\it p}$-waves.  

In this work, we further investigate the simulations reported in \citet{Fragile18} to study the diskoseismic oscillations. The initial configurations of these simulations are set using standard thin disk model from \citep{Shakura73}. Although these simulations are purely hydrodynamical, their study can shed some light on observed properties of high frequency QPOs. The numerical methods and initial setups (Table \ref{tab:params}) of the reported simulations are described in detail in \citet{Fragile18}. In Section \ref{sec:mode} we discuss the results and findings for both {\it non-rotating} and {\it rotating} black hole cases. Section \ref{sec:breathing} and \ref{sec:resonance} briefly describe breathing and resonant oscillations at a 3:2 frequency ratio. In Section \ref{sec:comparison}, we compare our findings with previous results. Finally, in Section \ref{sec:conclusion}, we summarize our conclusions. Throughout the article, the units of length and time are the gravitational radius, $r_\mathrm{g} = GM/c^2$, and time, $t_\mathrm{g} = GM/c^3$, respectively.  
\begin{table*}[]
\caption{Simulation Models and Parameters. The models are named by the initial mass accretion rate in units of $L_\mathrm{Edd}/c^2$; $r_\mathrm{min}$,  $r_\mathrm{max}$, $\theta_\mathrm{min}$ and $\theta_\mathrm{max}$ correspond to the radial and polar boundary limits; the $q$ parameter dictates how concentrated the grid is toward the disk midplane ($q =0.1$ is more concentrated than $q=0.3$); and $t_\mathrm{stop}$ represents the stop time of the numerical simulation.}
\centering                          
\begin{tabular*}{2\columnwidth}{@{\extracolsep{\fill}}  l   cccccccc}
\\
\hline \hline
\\
Name & BH Spin & $\dot{m}$ & $r_\mathrm{min}$ & $r_\mathrm{max}$ & $\theta_\mathrm{max} - \theta_\mathrm{min}$ & $q$ & $t_\mathrm{stop}$ \\
 &  & ($L_\mathrm{Edd}/c^2$) &($r_\mathrm{g}$) &($r_\mathrm{g}$) & (rad) &  &($t_\mathrm{g}$)
\\
\hline \hline
\\
S01E & 0.0 &0.01 & 5 & 20 & 0.289 & 0.1 & 80,485 \\
S01Ea5 & 0.5 & 0.01 & 4 & 20 & 0.289 & 0.1 & 80,485 \\
S1E & 0.0 & 1 & 5 & 20 & 0.401 & 0.1 &  80,652\\
S3E & 0.0 & 3 & 4 & 40 & 0.476 & 0.3 & 42,964 \\
S3Ep & 0.0 & 3 & 4 & 40 & 0.476 & 0.3 & 29,665 \\
S10E & 0.0 & 10 & 4 & 40 & 0.871 & 0.3 & 27,492
\end{tabular*}
\label{tab:params}
\end{table*}

\section{Disk Oscillations}
\label{sec:mode}

We now turn to the inferred oscillations from our set of thin disk simulations. Unlike previous simulations of thin, viscous accretion disks, the \citet{Fragile18} simulations were carried out in a GR framework with  radiative transfer. The radiation ws treated  using the M1-closure  scheme \citep{Mihalas84}. As described above, the presence of ${\it g}$-modes and ${\it p}$-modes in the inner part of a thin viscous disk has been investigated in Newtonian simulations with the Paczy\'nski-Wiita pseudo-potential \citep{ONeill09}. Analytic diskoseismic studies \citep{Nowak92,Perez97} reveal that the oscillatory modes of a thin accretion disk in the Schwarzschild metric are qualitatively similar to those obtained in a Newtonian analysis with a pseudo-potential in which the radial epicyclic frequency has a maximum and drops to zero at the ISCO. On these grounds, we expect that our simulations will reveal similar modes as the ones in \citet{ONeill09}. However, this is by no means certain, as their simulation had an effectively optically thin disk (entropy was artificially removed at the rate of viscous heating) while our simulation includes full radiative transfer, and the disk is optically thick. Thus, one aim of our study is to find out whether the inclusion of radiation suppresses or preserves the ${\it g}$- and ${\it p}$-modes reported in previous numerical, non-radiative, Newtonian studies. 

\subsection{Simulation S1E (Radiation-Pressure-Dominated Setup)}
\label{subsec:S1E}
\begin{figure*}
\centering
\includegraphics[width=\columnwidth]{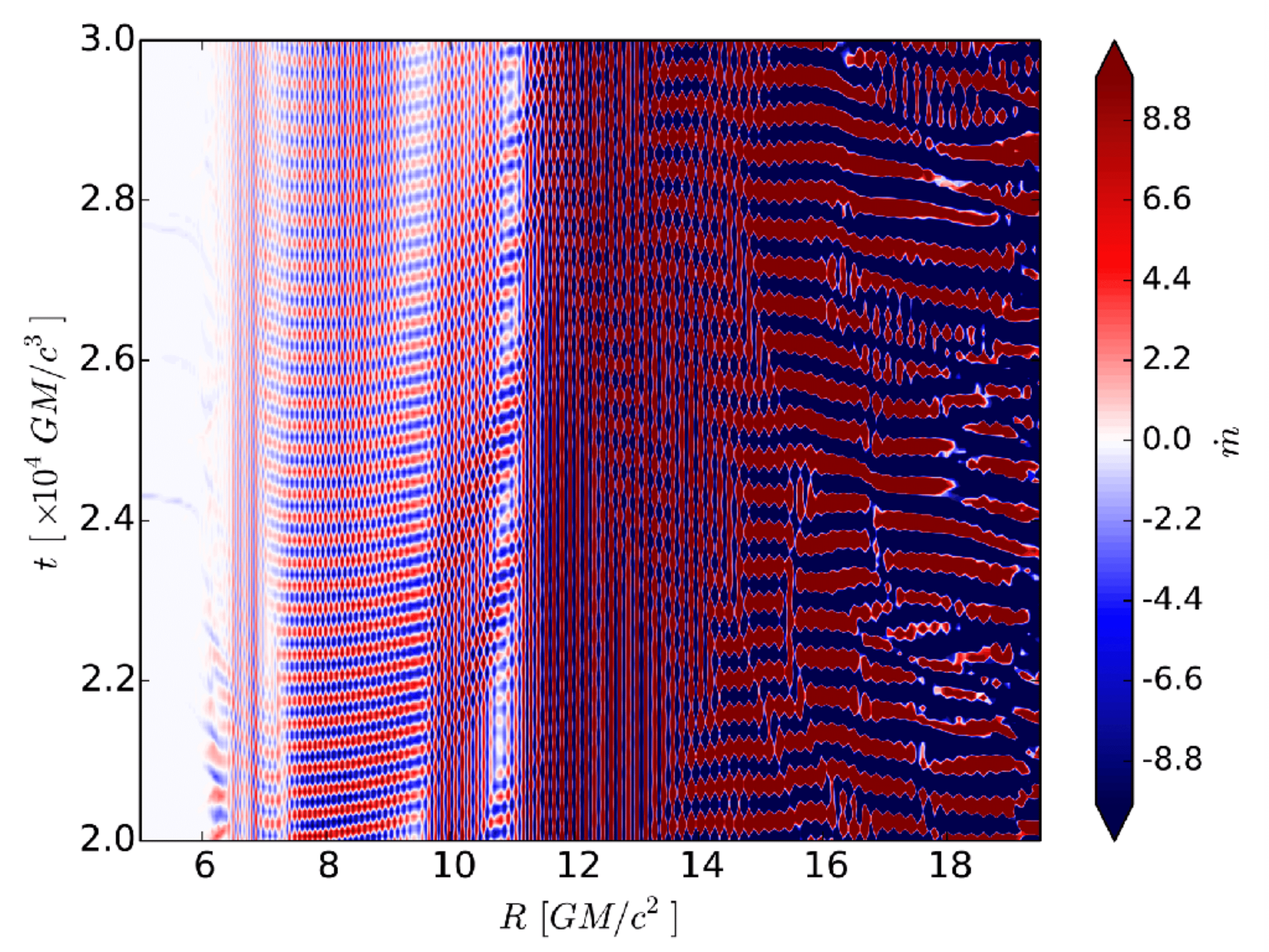}
\includegraphics[width=\columnwidth]{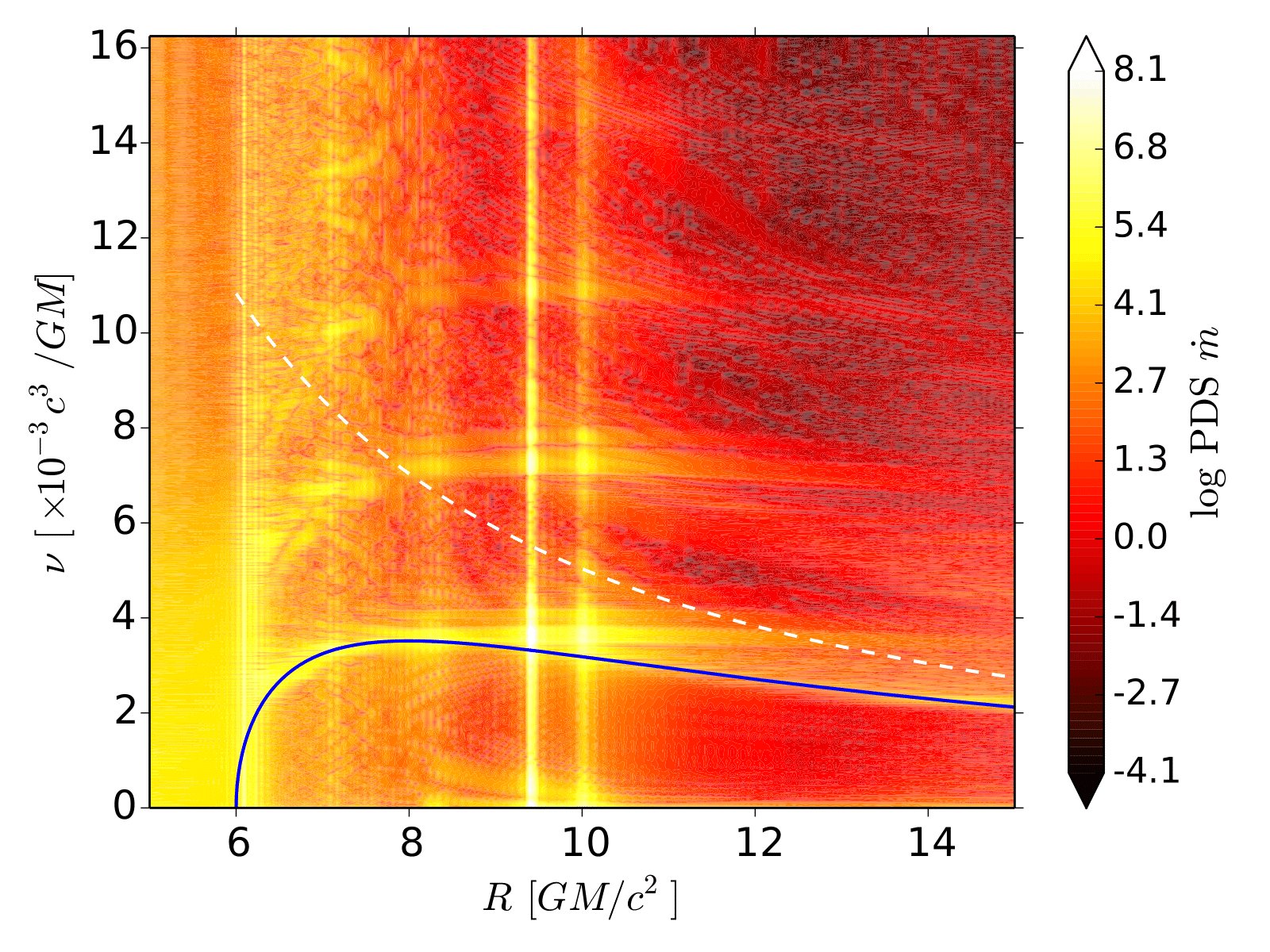}
\caption{Space-time plot (left) and PDS (right) of the local mass accretion rate over the complete duration ($80\,652\,t_\mathrm{g}$) of simulation S1E. The solid blue line in the right panel, and in analogous figures below, is a plot of the radial epicyclic frequency $\kappa(r)/(2\pi)$; the white dashed line is the Keplerian frequency $\Omega(r)/(2\pi)$.
\label{fig:dsfig1}}
\end{figure*}

The most robust timing feature in our simulations is a persistent, coherent oscillation in the local mass accretion rate $\dot{m}(r)$ (mass flux through a sphere of radius $r$). This is qualitatively seen in each of the reported simulations. An example is given in the left panel of Fig.~\ref{fig:dsfig1} for the S1E simulation in the time interval $2.0$--$3.0 \times 10^4\,t_\mathrm{g}$. The oscillation at $r\approx 8\,r_\mathrm{g}$ is quite coherent, persisting throughout the simulation. The reader can easily count 35 nearly horizontal stripes/cycles in the exhibited time interval of $10^4\,t_\mathrm{g}$, yielding a frequency $\nu =3.5\times 10^{-3}\,t_\mathrm{g}^{-1}$. Note that in the upper half of the diagram ($t > 2.5 \times 10^4\,t_\mathrm{g}$), the stripes are horizontal at $r \approx 8\,r_\mathrm{g}$, and are curving upwards on both sides, consistent with a trapped oscillation exciting traveling waves on both sides of the trapping domain. On the other hand, at the bottom of the figure ($t < 2.3 \times 10^4\,t_\mathrm{g}$), the stripes are consistent with a wave starting at $r\approx 7.5\,r_\mathrm{g}$ traveling towards larger radii.

The corresponding power density spectrum (PDS) of mass flux in the right panel of Fig.~\ref{fig:dsfig1} shows the power in a fast Fourier transform (FFT) of the entire time series from $t=0$ to $80\,652\,t_\mathrm{g}$. The power is normalized by the zeroth frequency power (which is the initial mass accretion rate in the case of Fig.~\ref{fig:dsfig1}). There is remarkably little power below the radial epicyclic frequency curve $\tilde{\kappa}(r)$ (the solid blue curve), in agreement with the predictions of the linear theory of diskoseismology,
which only allows for ${\it g}$-modes in this region, and only at a frequency close to $\tilde{\kappa}_\mathrm{max}$.

\begin{figure*}
\includegraphics[width=\columnwidth]{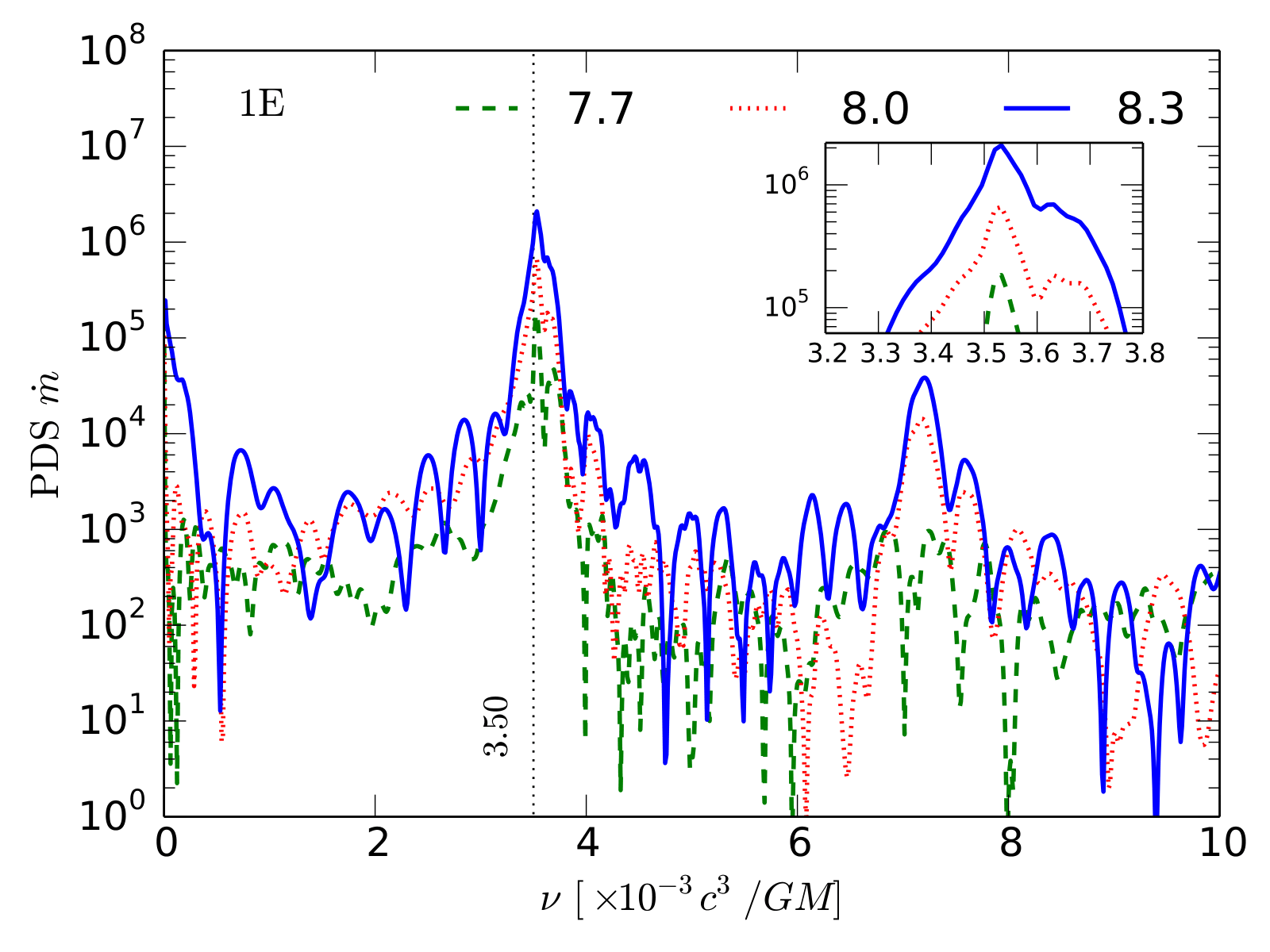}
\includegraphics[width=\columnwidth]{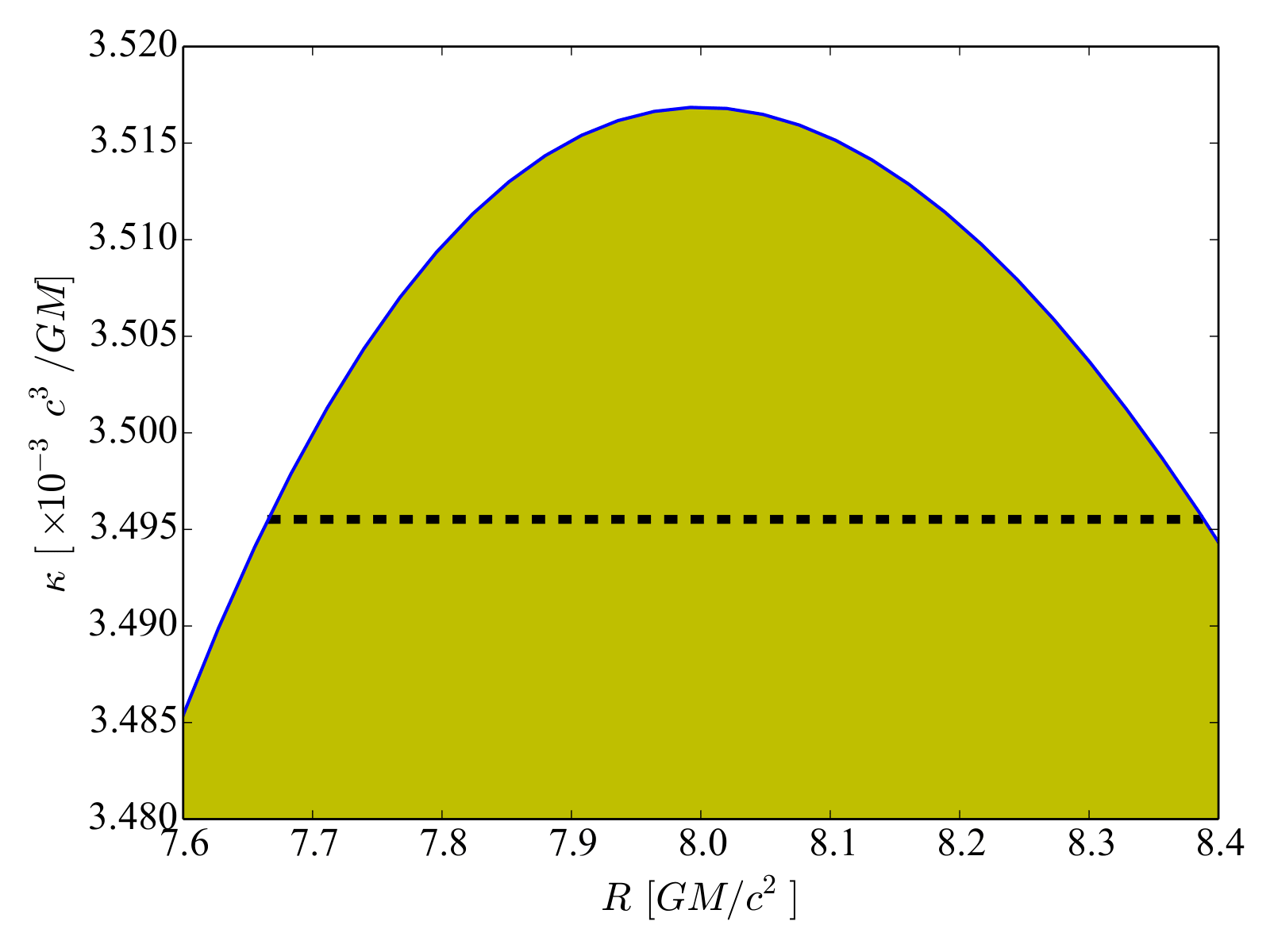}
\caption{Left: PDS of local mass accretion rate at three radii ($7.7$, $8.0$ and $8.3\,r_\mathrm{g}$) for simulation S1E. Right: Lowest-frequency, trapped {\it g}-mode frequency (dashed line) over the trapping region defined by the radial epicyclic frequency (blue, solid line) for the Schwarzschild metric.
\label{fig:dsfig2}}
\end{figure*}

\begin{figure*}
\includegraphics[width=\columnwidth]{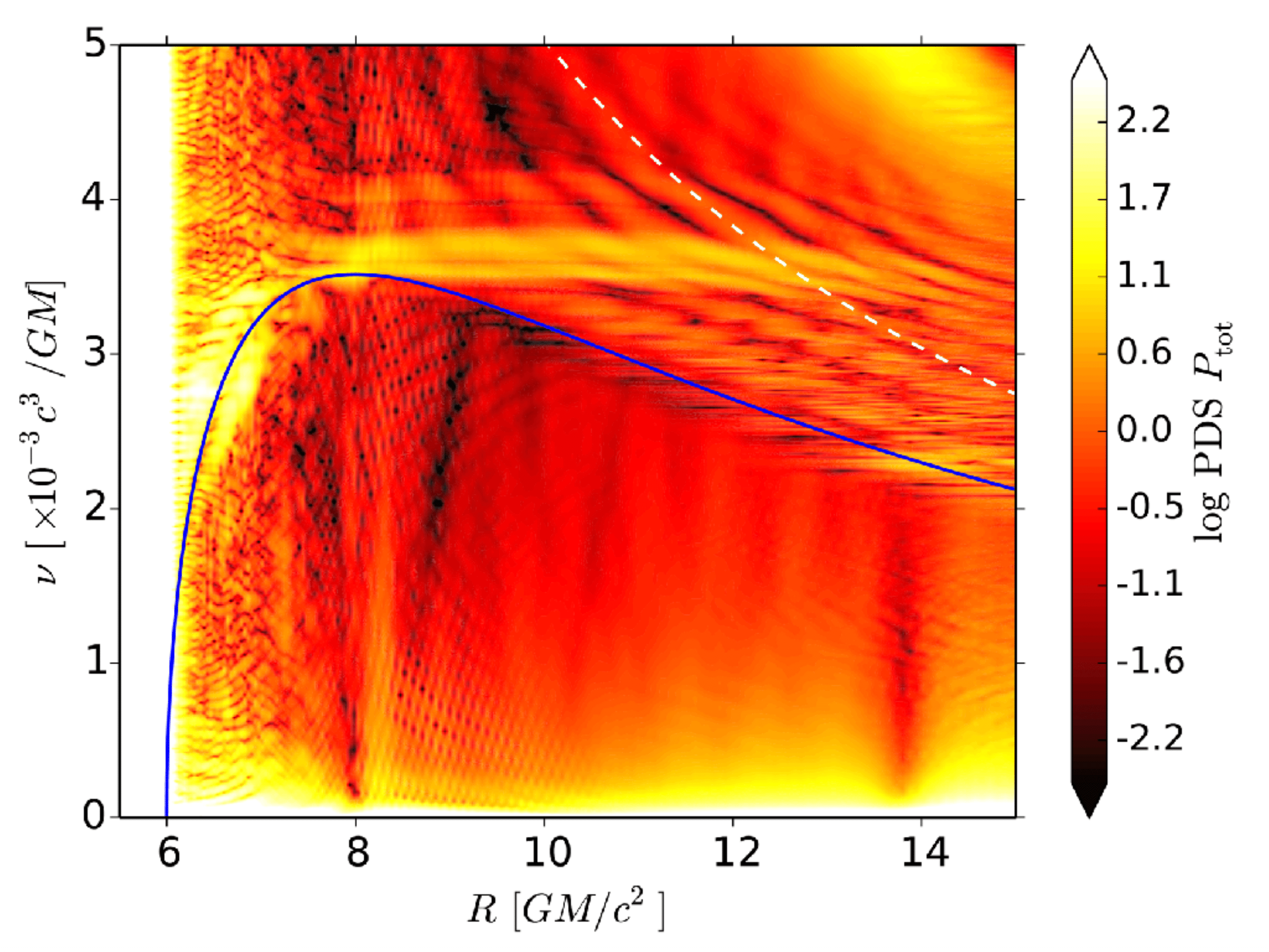}
\includegraphics[width=\columnwidth]{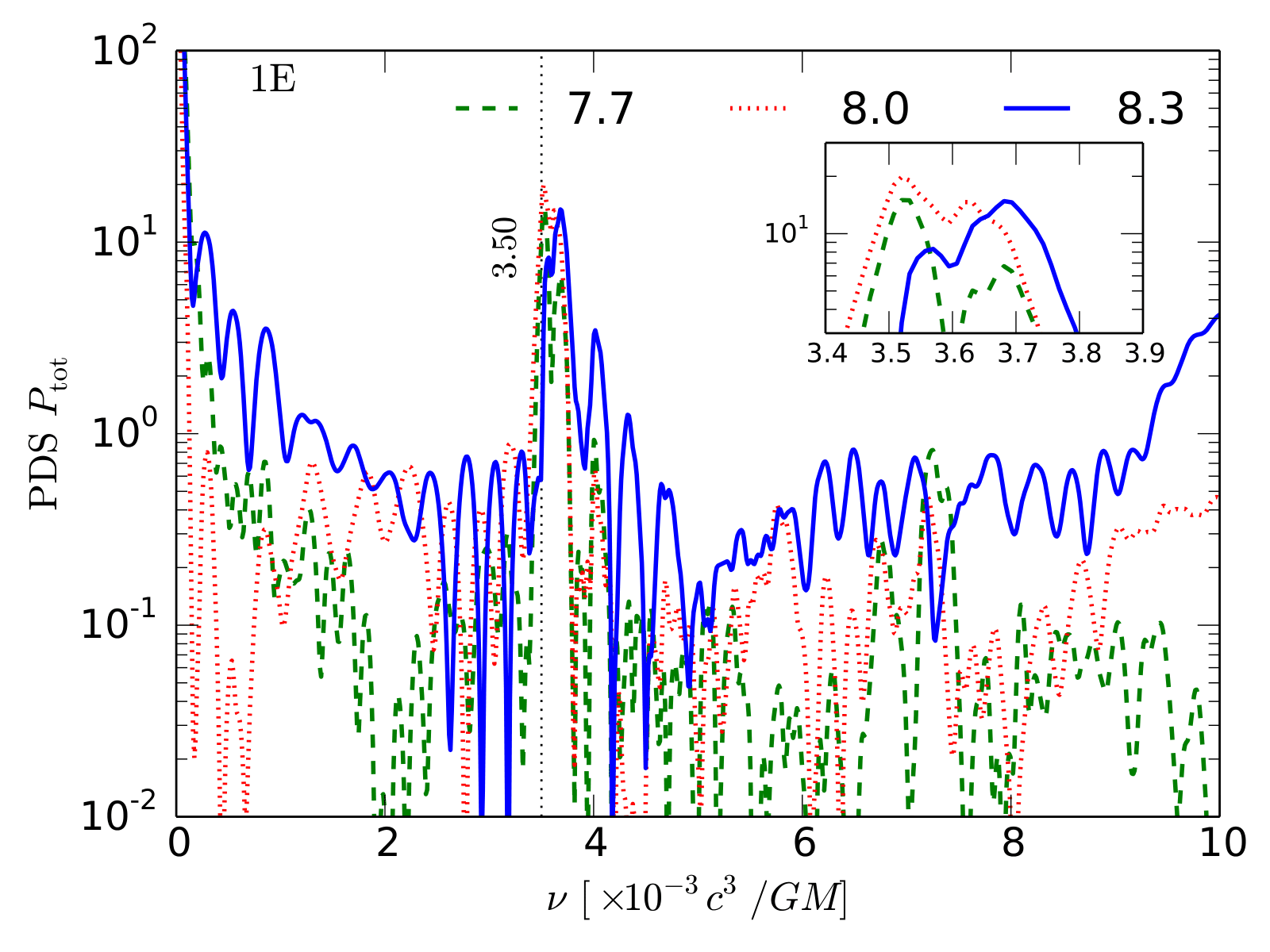}
\caption{Left: PDS of total midplane pressure for simulation S1E. Right: Same at radii 7.7, 8.0, and $8.3\,r_\mathrm{g}$. The vertical dotted line shows the lowest eigenfrequency for a trapped {\it g}-mode.
\label{fig:dsfig3}}
\end{figure*}

Ideally, we aim to identify the frequency peak and classify it to see if it belongs to the set of theoretically calculated diskoseismic modes of oscillation. As discussed earlier, the maximum epicyclic frequency in the Schwarzschild metric is $\tilde{\kappa}= 3.51686\times 10^{-3}\,c^3/(GM)$, while the lowest-frequency  ${\it g}$-mode in a $0.1\,L_\mathrm{Edd}$ disk has the frequency $ 3.35723\times 10^{-3}\,c^3/(GM)$ and  the lowest-order fundamental ${\it g}$-mode has $\nu_0= 3.49552\times 10^{-3}\,c^3/(GM)$ \citep{Perez97}, the difference being only $1.6\times 10^{-4}\,c^3/(GM)$ or $2\times 10^{-5}\,c^3/(GM)$, respectively. It would, therefore, be difficult to resolve these frequencies in a simulation run for a time of $<10^5\,t_\mathrm{g}$. The relation between these two frequencies is illustrated in the right panel of Fig.~\ref{fig:dsfig2}, the dashed line showing the radial extent of the trapping region of the fundamental ${\it g}$-mode. As the ${\it g}$-mode becomes evanescent on each side of this radial domain, the amplitude of the oscillation close to the boundaries of the trapping range should be lower than that at the center of the trapped domain ($r=8\,r_\mathrm{g}$ in the Schwarzschild spacetime). This does not seem to be the case in any of  the viscous runs of \citet{ONeill09}, although their inviscid run may have this property (their Fig. 2). We also do not see the amplitude behaving this way in the PDS of the mass flux (although, we see a hint of this in the PDS of pressure for the S1E run, as seen in the right panel of Fig.~\ref{fig:dsfig3}). We suspect that such a behavior of the amplitude for ${\it p}$- and ${\it g}$-modes is due to their mutual interference.

The left panel of Fig.~\ref{fig:dsfig2} shows the {\it local} PDS of $\dot{m}(r)$ for the same S1E run computed at $r=7.7$,  $8.0$, and $8.3\,r_\mathrm{g}$ (i.e., three vertical slices through the PDS in the right panel of Fig.~\ref{fig:dsfig1}). The most prominent peak is at least a factor of $100$ above the background noise and is at a frequency very close to $3.52\times 10^{-3}\,c^3/(GM)$. There are four possibilities for the oscillation type at this frequency: 1. the fundamental ${\it g}$-mode; 2. radial oscillations at the maximum radial epicyclic frequency; 3. ${\it p}$-modes leaking into the evanescent region; and 4. a blend of the $g$-mode and penetrating ${\it p}$-modes. Note that the peak power at $r=8.3\,r_\mathrm{g}$ exceeds that at $r=8\,r_\mathrm{g}$ by a factor of about 3, disfavoring the expectations of a ${\it p}$-mode originating to the left of the $\tilde{\kappa}$ curve ($r<8\,r_\mathrm{g}$) and leaking/exciting a ${\it g}$-mode. If this were the case, the maximum power should have been at $r=7.7\,r_\mathrm{g}$. A ${\it p}$-wave originating  to the right of the $\tilde{\kappa}$ curve ($r>8\,r_\mathrm{g}$) can also leak into the evanescent region and enhance the power at $r=8.3\,r_\mathrm{g}$. Despite the fact that the power in the left panel of Fig. \ref{fig:dsfig2} is not maximal at $r=8\,r_\mathrm{g}$, the presence of a ${\it g}$-mode  cannot be discarded. We argue this on the basis that ${\it p}$-modes leaking inside the evanescent region can enhance the power in the ${\it g}$-mode near the edges of the trapping region (shown by the dashed black line in the right panel of Fig.~\ref{fig:dsfig2}).

A further and better understanding of the ${\it g}$-mode in our simulations can be achieved by examining the power density spectrum of total pressure variations in the S1E disk (Fig.~\ref{fig:dsfig3}). Here, the most prominent feature is a stripe covering nearly the complete radial range of the simulation and corresponding to a frequency just slightly larger than $\tilde{\kappa}_\mathrm{max}$, although there is also some power just below $\tilde{\kappa}_\mathrm{max}$. The right panel of Fig.~\ref{fig:dsfig3} again shows the {\it local} PDS of total pressure. The PDS are computed at the same radii as before. No harmonics are apparent in the PDS of total pressure above the noise level, contrasting with the PDS of $\dot{m}$ in Figs.~\ref{fig:dsfig1} and \ref{fig:dsfig2}. A closer examination (inset) reveals two frequency peaks for each radius, while for two of the radii ($7.7$ and $8.0$), one of the peaks is precisely at $3.516\times 10^{-3}\,t_\mathrm{g}^{-1}$, i.e., at $\tilde{\kappa}_\mathrm{max}$. The extra power at $3.7\times 10^{-3}\,t_\mathrm{g}^{-1}$ in the blue curve indicates the effects of the ${\it p}$-mode having $\nu >\tilde{ \kappa}_\mathrm{max}$. The red curve has its maximum exactly at the ${\it g}$-mode frequency as expected. This suggests that the ${\it g}$-mode is present irrespective of ${\it p}$-modes of slightly different frequencies ($\nu > \tilde{\kappa}_\mathrm{max}$) altering/blending the power in the region with $\nu < \tilde{\kappa}_\mathrm{max}$. The frequency-radius PDS diagram in the left panel of Fig. \ref{fig:dsfig3} also shows ${\it p}$-waves at the frequency of the fundamental ${\it g}$-mode, in addition to standing ${\it p}$-waves at slightly higher frequencies (note the narrow, diagonal, dark strips strongly suggestive of lines of nodes).

\subsection{Simulation S01E (Gas-Pressure-Dominated Setup)}
\label{subsec:S01E}
\begin{figure*}
\includegraphics[width=\columnwidth]{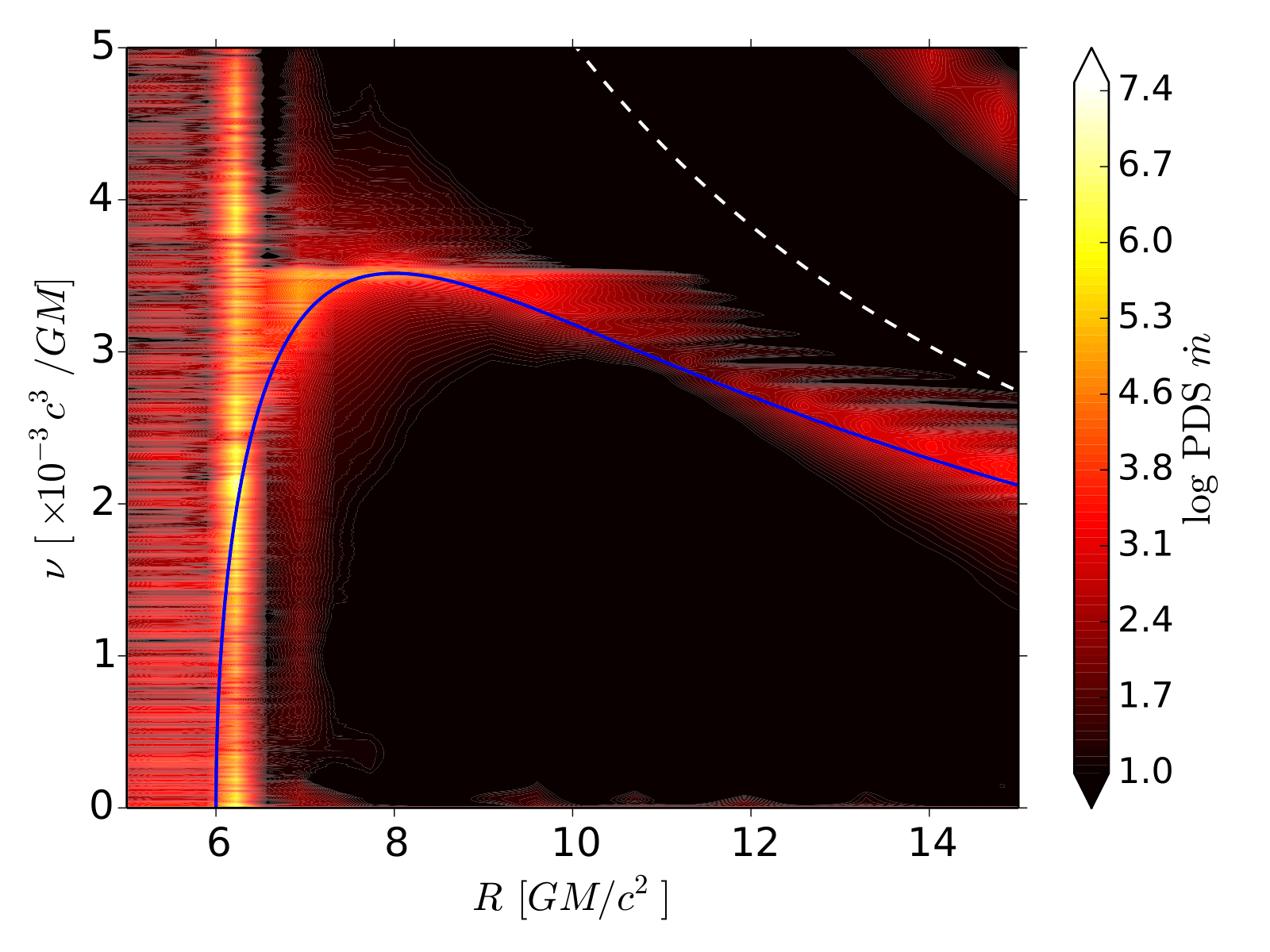}
\includegraphics[width=\columnwidth]{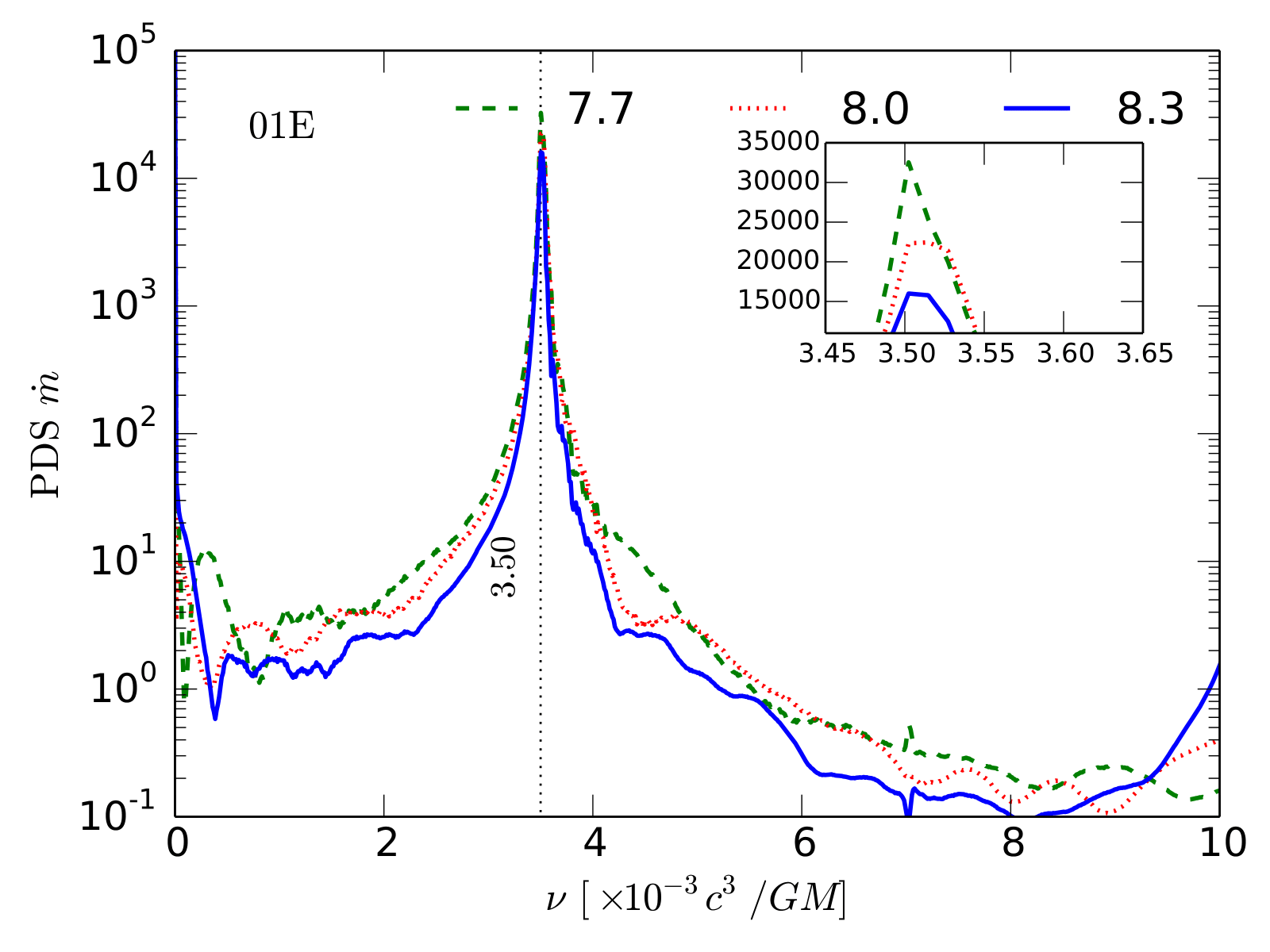}
\caption{Left: PDS of the local mass accretion rate for simulation S01E. Right: Local PDS at radii 7.7, 8.0, and $8.3\,r_\mathrm{g}$. The vertical dotted line represents the lowest trapped ${\it g}$-mode eigenfrequency.
\label{fig:dsfig4}}
\end{figure*}

Up to this point we have been discussing a simulation that was initialized assuming a radiation-pressure-dominated setup, but actually achieved a gas-pressure-dominated equilibrium following its thermal collapse \citep[see Fig. 1 in][]{Fragile18}. Let us now turn to the one case of a thermally stable disk from \citet{Fragile18}, that being simulation S01E. Similar to simulation S1E, this disk manifests strong, persistent and coherent oscillations in its local mass flux. The left panel of Fig.~\ref{fig:dsfig4} shows the PDS of $\dot m(r)$ performed over the entire duration ($80 485\,t_\mathrm{g}$) of simulation S01E. The disk spectrum is remarkably quiet. The white noise near the ISCO dominates the power. Other than a feature near $\tilde{\kappa}_\mathrm{max}$, there is hardly any power at all below the radial epicyclic frequency curve. However, there is excess power at radii $r<8\,r_\mathrm{g}$ and frequencies in the range $\tilde{\kappa}(r)<\nu(r)<\tilde{\kappa}_\mathrm{max}$. As predicted in linear theory \citep{Kato2001} these oscillations correspond to inner trapped ${\it p}$-modes between the ISCO and the epicyclic barrier. With a finer grid one could possibly resolve the number of radial nodes in such oscillations, and with a longer run resolve the discrete spectrum of these modes. With the simulations performed to date, this is not feasible.

Similar to simulation S1E, there is power to the right of the radial epicyclic curve (i.e., for $r>8\,r_\mathrm{g}$). These oscillations correspond to ${\it p}$-modes excited by epicyclic oscillations and extend to the outer boundary of the computational domain at $r=20\,r_\mathrm{g}$ (for simulation S01E). Alternatively, they could also consist of $n=1$ (single vertical node) ${\it g}$-modes, which are trapped in the range $\tilde{\kappa}<\nu<\tilde N_z$, where $\tilde N_z$ is the Brunt-V\"ais\"all\"a frequency, which for our strongly stratified disk is expected to be on the order of the vertical epicyclic frequency, which is equal to the Keplerian orbital frequency (white dashed line in the figure) in the Schwarzschild metric, so $\tilde N_z \sim \tilde\Omega$. It is remarkable that there are hardly any oscillations at frequencies $\nu>\tilde{\kappa}_\mathrm{max}$. The strip of power in the upper right corner of the  left panel of Fig.~\ref{fig:dsfig4} is tied to the breathing oscillation \citep{Mishra18}, discussed in Section ~\ref{sec:breathing}.

The most prominent feature in Fig. \ref{fig:dsfig4}, though, is the oscillation at $\nu=\tilde{\kappa}_\mathrm{max}$, extending from the inner edge of the disk to $r\approx 12\,r_\mathrm{g}$. This is seen as a rather narrow peak at $\nu=3.50\times 10^{-3}\,t_\mathrm{g}^{-1}$ in the right panel of Fig.~\ref{fig:dsfig4}, which shows the PDS at three distinct local radii, $r=7.7$, 8.0, and $8.3\,r_\mathrm{g}$. Note that the peak power is at least three orders of magnitude above the background power, and the harmonic is prominently absent (relative to the peak, the power is down by four or five orders of magnitude at $2\tilde{\kappa}_\mathrm{max}$). Again, there is less power in the $\nu=3.5\times 10^{-3}\,t_\mathrm{g}^{-1}$ peak at $r=8\,r_\mathrm{g}$, than at one of the radii at the edge of the trapping zone of the ${\it g}$-mode, this time at the inner edge at $r=7.7\,r_\mathrm{g}$. This advocates a leakage of ${\it p}$-mode into the evanescent region and constructive coupling to the ${\it g}$-mode, leading to larger power in the green curve. We noted a similar feature in the S1E case but from a larger radius.

\begin{figure*}
\includegraphics[width=\columnwidth]{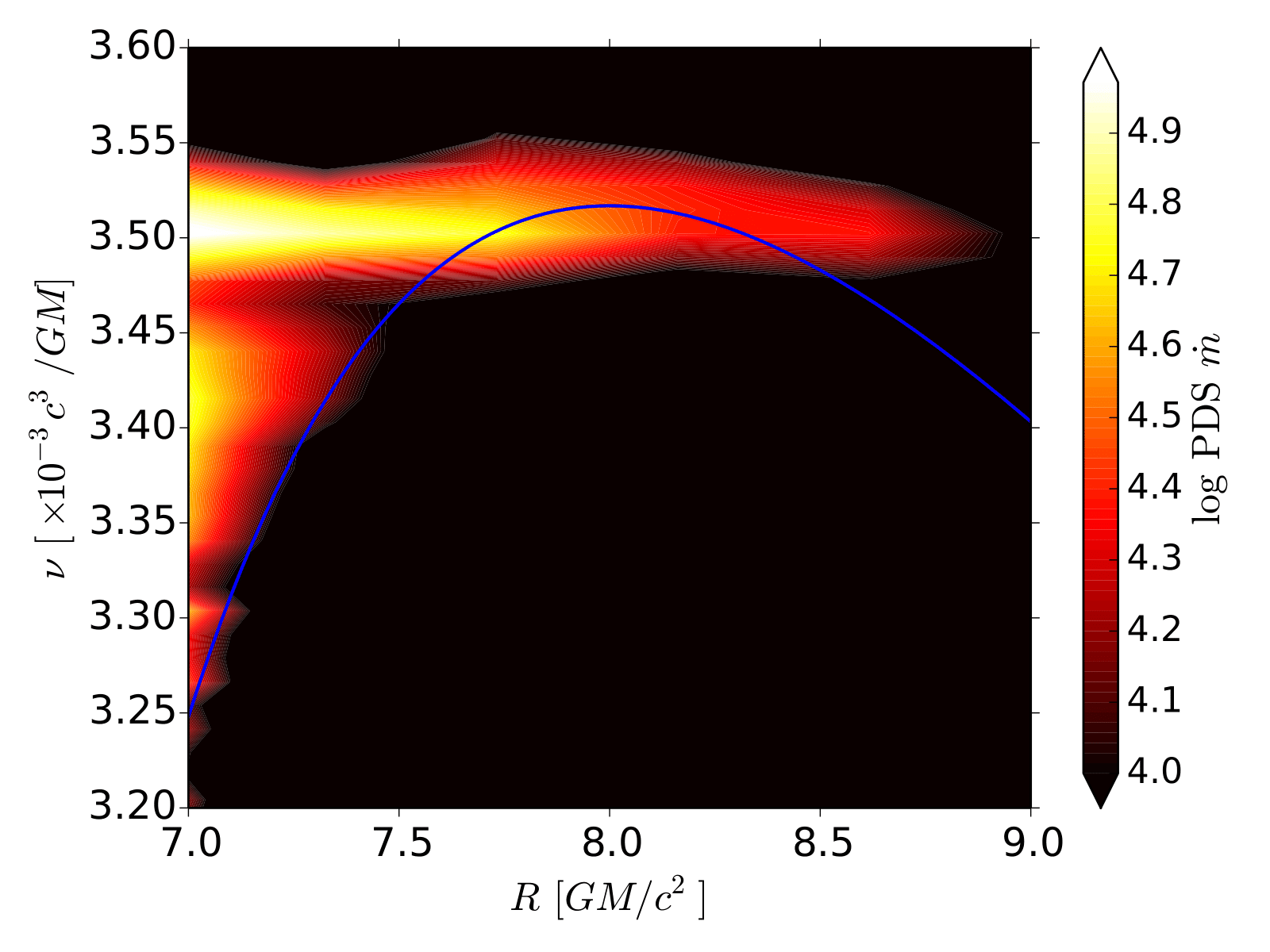}
\includegraphics[width=\columnwidth]{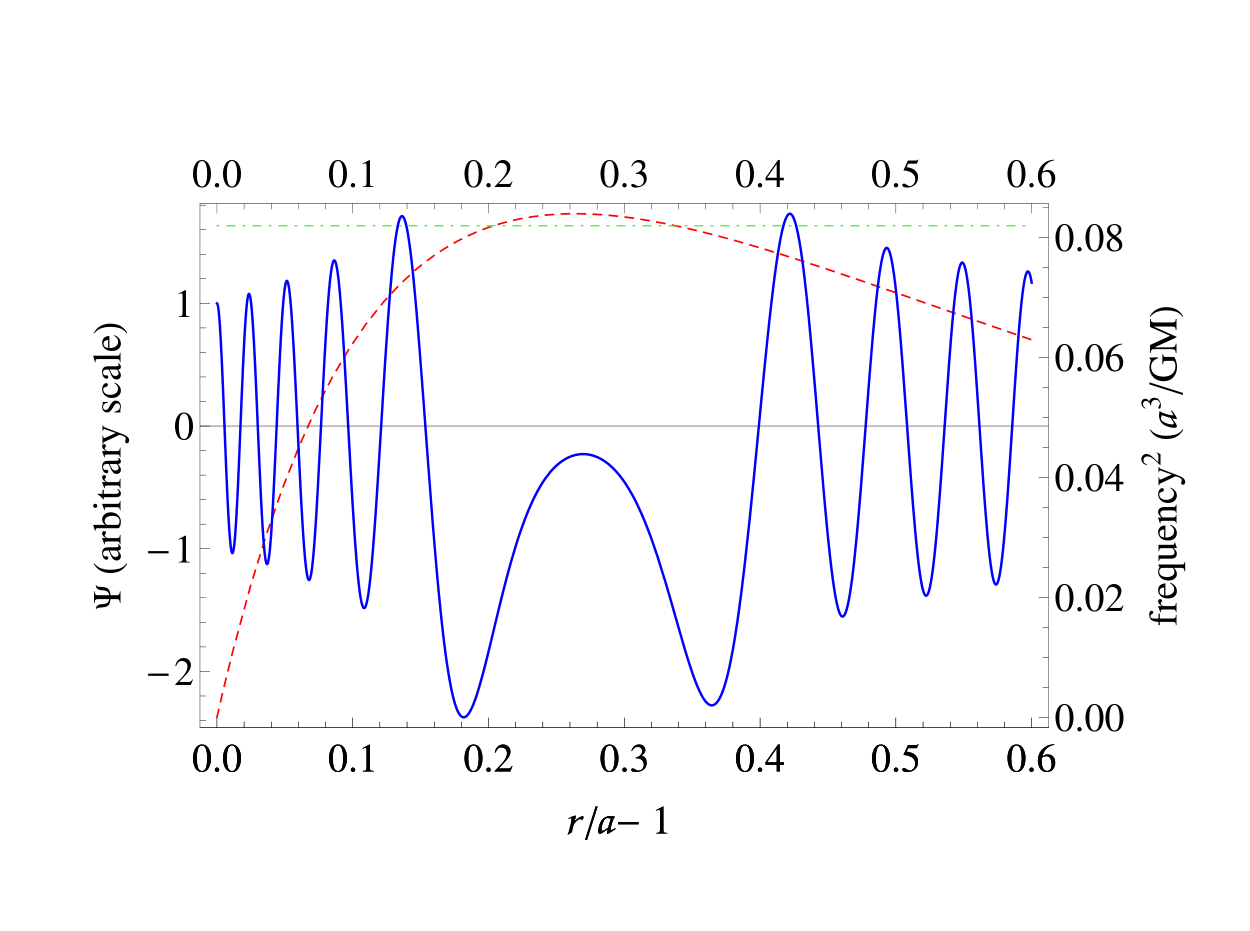}
\caption{Left: Zoomed PDS for the local mass accretion rate of simulation S01E. Right: A {\it p}-mode (solid, blue line) from the continuum part of the spectrum transmitted through the evanescent region identified by the intersection of $\kappa(r)$ (dashed, red line) and the eigenfrequency (dot-dashed, green line). Here, $a$ is the radius of the inner edge of the disk \citep[plot reproduced from][]{Giussani14}.
\label{fig:dsfig5}}
\end{figure*}

Alternatively, we could try to interpret the strong oscillations seen at three different radii in the right panel of Fig.~\ref{fig:dsfig4} as a ${\it p}$-mode transmitted through the evanescent region, where $\nu<\tilde{\kappa}(r)$. Such a possibility follows from the work of \citet{Giussani14}, who computed the entire spectrum of the axisymmetric ($m=0$) inertial-acoustic modes in the pseudo-potential of \citet{KL2002}, and showed that the modes with up to eight nodes belong to the discrete spectrum (i.e., they are trapped between the ISCO and the $\tilde{\kappa}(r)$ curve), while standing waves with nine radial nodes in the inner trapping region are transmitted through the evanescent region to larger radii. Also, the lower boundary of the continuum frequency spectrum of ${\it p}$-waves is below $\tilde{\kappa}_\mathrm{max}$. This is illustrated in the right panel of Fig.~\ref{fig:dsfig5}, reproduced from \citet{Giussani14}, showing one such eigenmode. The plot clearly shows that, based on this hypothesis, the power at $r=8\,r_\mathrm{g}$ should be less than at both edges of the evanescent region. Again, this is not what we see in our simulations, as the left panel of Fig.~\ref{fig:dsfig5} clearly reveals a decline of power with increasing radius. However, the high $\alpha$ results of \citet{ONeill09} do show more power at the edges of the evanescent zone than at the radial location of $\tilde{\kappa}_\mathrm{max}$ in their vertical velocity plots, with the exception of model EQ0.1 (their Figs. 7 and 8), suggesting that the transmission of ${\it p}$-waves with frequency less than $\tilde{\kappa}_\mathrm{max}$ is a possible alternative explanation to their ${\it g}$-mode feature (unless, of course, this is a ${\it g}$-mode with one radial node, which would also explain the lower power in the middle of the trapped domain).

%

\begin{figure*}
\includegraphics[width=\columnwidth]{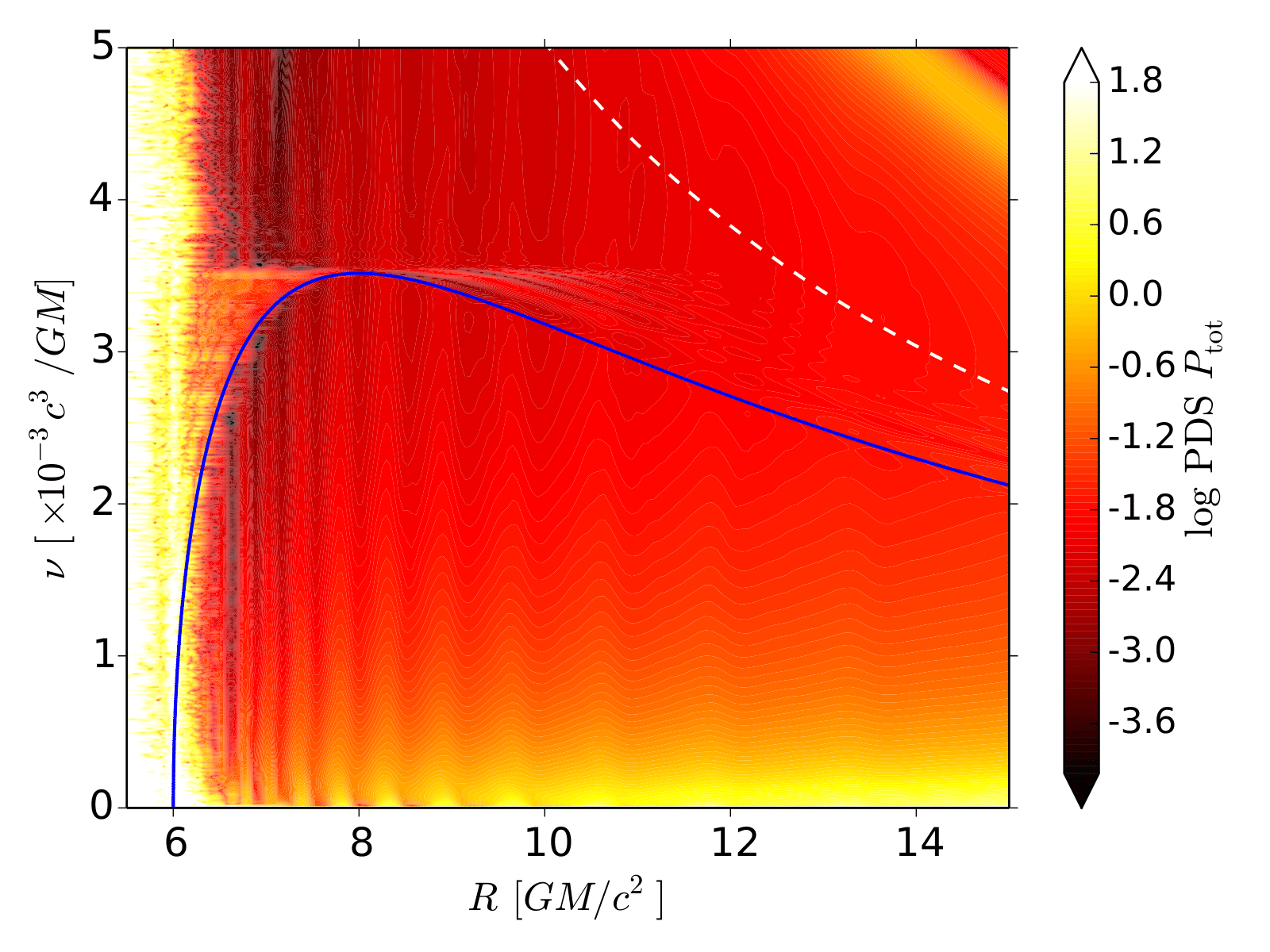}
\includegraphics[width=\columnwidth]{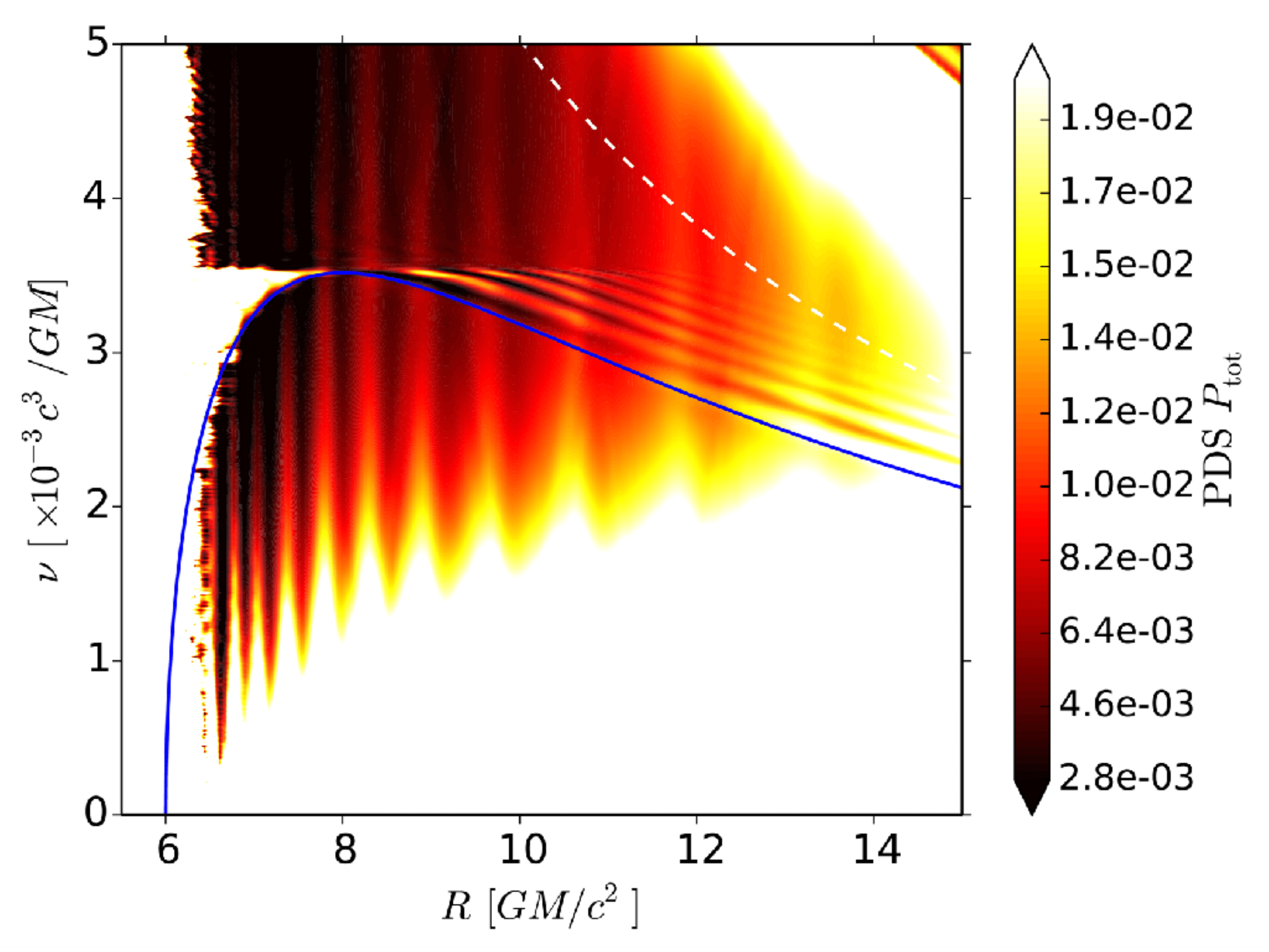}
\caption{Left: PDS for the total midplane pressure of simulation S01E. Right: Same in a restricted power range to enhance the the features at $r>8\,r_\mathrm{g}$ and $\nu>\kappa(r)/(2\pi)$. Note the lack of power for $\nu>\kappa_\mathrm{max}/(2\pi)$ in the range $6<r<12$.
\label{fig:dsfig6}}
\end{figure*}

Considering the PDS of total (gas + radiation) pressure for the same S01E simulation, as shown in Fig.~\ref{fig:dsfig6}, we draw similar conclusions as with the PDS of $\dot{m}$. Again, there is remarkably little noise in the disk (except at very low frequencies). Just as in the left panel of Fig.~\ref{fig:dsfig4}, there is clearly a fair amount of power in the domain expected for trapped ${\it p}$-modes, at radii $r<8\,r_\mathrm{g}$ and frequencies $\tilde{\kappa}(r)<\nu<\tilde{\kappa}_\mathrm{max}$, an excess of power at $\nu=\tilde{\kappa}_\mathrm{max}$ extending from the ISCO to $r \approx 12\,r_\mathrm{g}$, and some power to the right of the epicyclic curve, i.e., at radii larger than those radii for which $\nu=\kappa(r)$, with $\nu< \tilde{\kappa}_\mathrm{max}$. Note the alternating dark and bright, diagonal stripes above the radial epicyclic frequency curve for $r>8\,r_\mathrm{g}$; the right panel of the figure has a restricted power range to enhance these stripes. The dark lines are the lines of nodes of oscillations. At a fixed frequency (horizontal line in the figure) one can clearly see alternating nodes and crests/troughs of the standing wave. The wavelength decreases with increasing radius. This argues for fundamental $(n=0)$ ${\it p}$-modes, rather than $n=1$ ${\it g}$-modes trapped by the Brunt-V\"ais\"all\"a frequency, as the wavenumber drops to zero at the edges of the evanescent zone. For the $n=0$ ${\it p}$-modes there is only one edge, at that radius $r_\mathrm{in}$ at which $\nu=\tilde{\kappa}(r_\mathrm{in})$, while for the $n=1$ ${\it g}$-modes there would be a second (outer) boundary at $r_\mathrm{out}$, with $\nu(r_\mathrm{out})=\tilde N_z(r_\mathrm{out})\approx \tilde\Omega(r_\mathrm{out})$, the latter frequency being given by the dashed white line. The vertical stripes are due to numerical noise present at all radii. However, these vertical stripes have very low power at the higher frequencies that we are interested in in this study and hence the conclusions of this article are not affected by these vertical stripes.

~~~~~~~~~~~

\subsection{Spinning black hole case}
\label{sec:spin}

In previous subsections, we quantitatively studied ${\it g}$-/${\it p}$-modes in accretion disks around {\it non-rotating} black holes. We also mentioned previously that the lowest order ${\it g}$-mode in the Schwarzschild case differs by only a fraction of a percent from the maximum radial epicyclic frequency. This makes it difficult to differentiate between oscillations at the maximum radial epicyclic frequency and ${\it g}$-mode frequency. In order to break this degeneracy, we performed a new S01E, gas-pressure-dominated disk simulation around a {\it rotating} black hole with dimensionless black hole spin parameter $a_* = 0.5$ (named S01Ea5).

\begin{figure*}
\begin{center}
\includegraphics[width=\columnwidth]{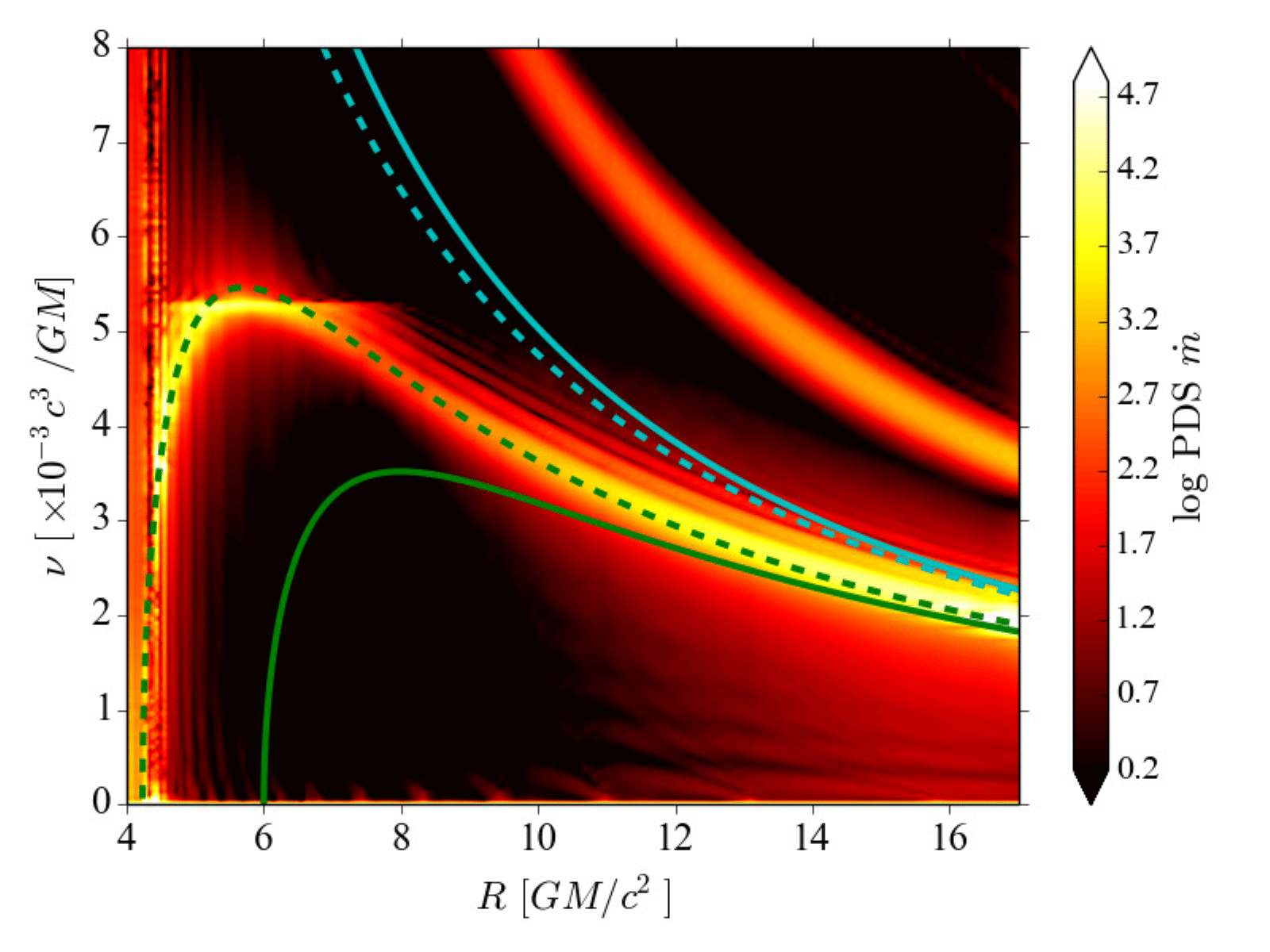}
\includegraphics[width=\columnwidth]{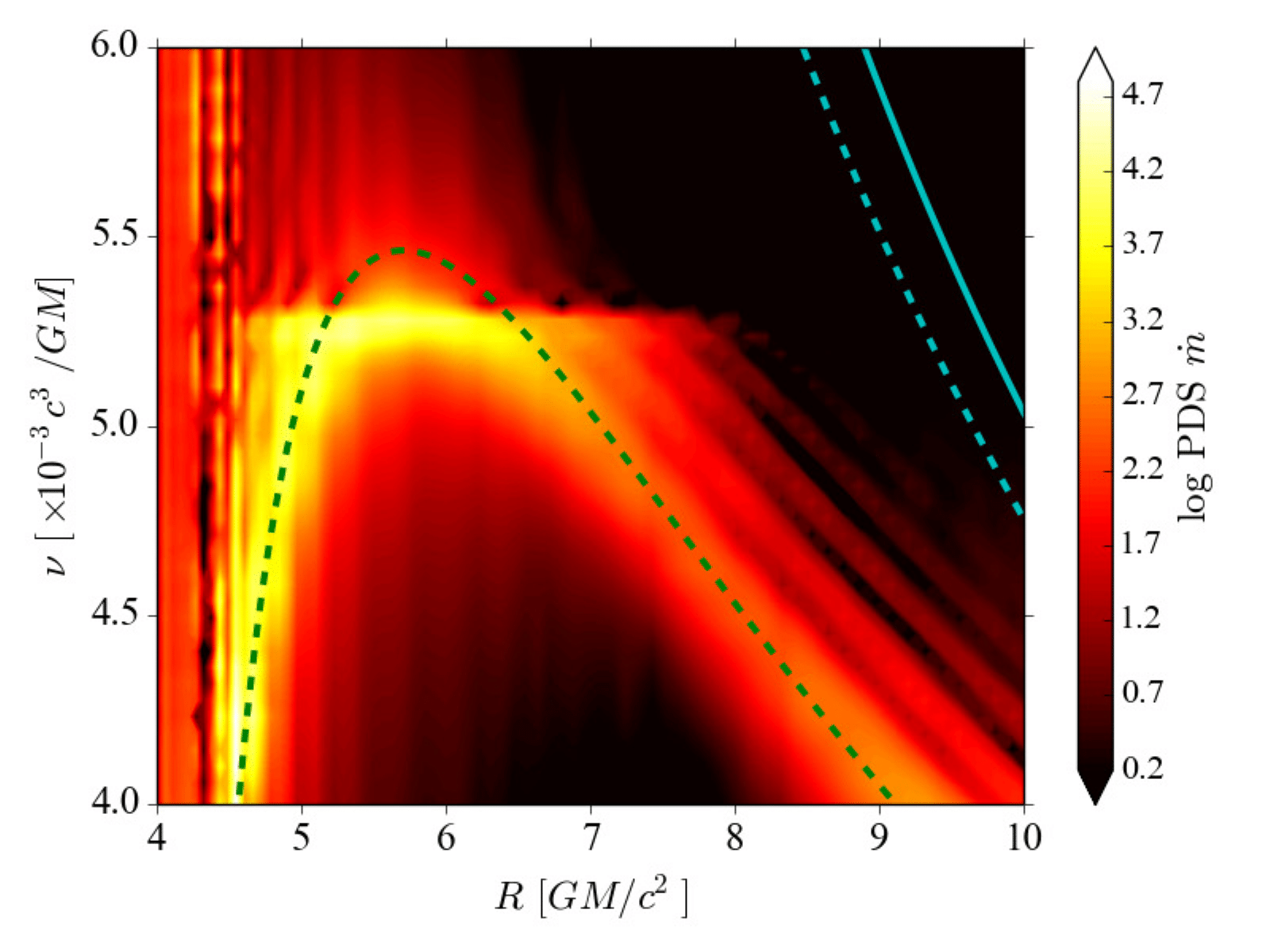}%
\caption{Left: PDS of the local mass accretion rate for simulation S01Ea5. Right: Zoomed in region of left panel. The solid curves show epicyclic frequencies for a non-rotating black hole, while the dashed curves are for an $a_*=0.5$ rotating black hole.}
\label{fig:bhspin}
\end{center}
\end{figure*}

In Fig.~\ref{fig:bhspin}, we show the radial profile of the PDS of $\dot{m}$. The left panel shows the global PDS, with solid curves representing the epicyclic frequencies for a non-rotating black hole and dashed curves representing the epicyclic frequencies for an $a_*=0.5$ black hole. The right panel magnifies the region close to the maximum radial epicyclic frequency ($\tilde{\kappa}_\mathrm{max} = 5.46\times 10^{-3}\,t_\mathrm{g}^{-1}$) to illustrate that most of the power is in a strip at $\nu= 5.28\times 10^{-3}\,t_\mathrm{g}^{-1}$. 
This strip is fairly wide in frequency, with a full width half maximum stretching roughly from $\nu= 5.23\times 10^{-3}\,t_\mathrm{g}^{-1}$ to $\nu= 5.3\times 10^{-3}\,t_\mathrm{g}^{-1}$ (Fig.~\ref{fig:local_a0p5}). These frequencies are fully compatible with the fundamental ${\it g}$-mode frequency $\nu_0=5.2315\times 10^{-3}\,t_\mathrm{g}^{-1}$ of a thin disk for an $a=0.5$ spin Kerr black hole, or another ${\it g}$-mode with no azimuthal or radial nodes at  $\nu=5.25\times 10^{-3}\,t_\mathrm{g}^{-1}$ \citep{Perez97}. 

\begin{figure*}
\begin{center}
\includegraphics[width=\columnwidth]{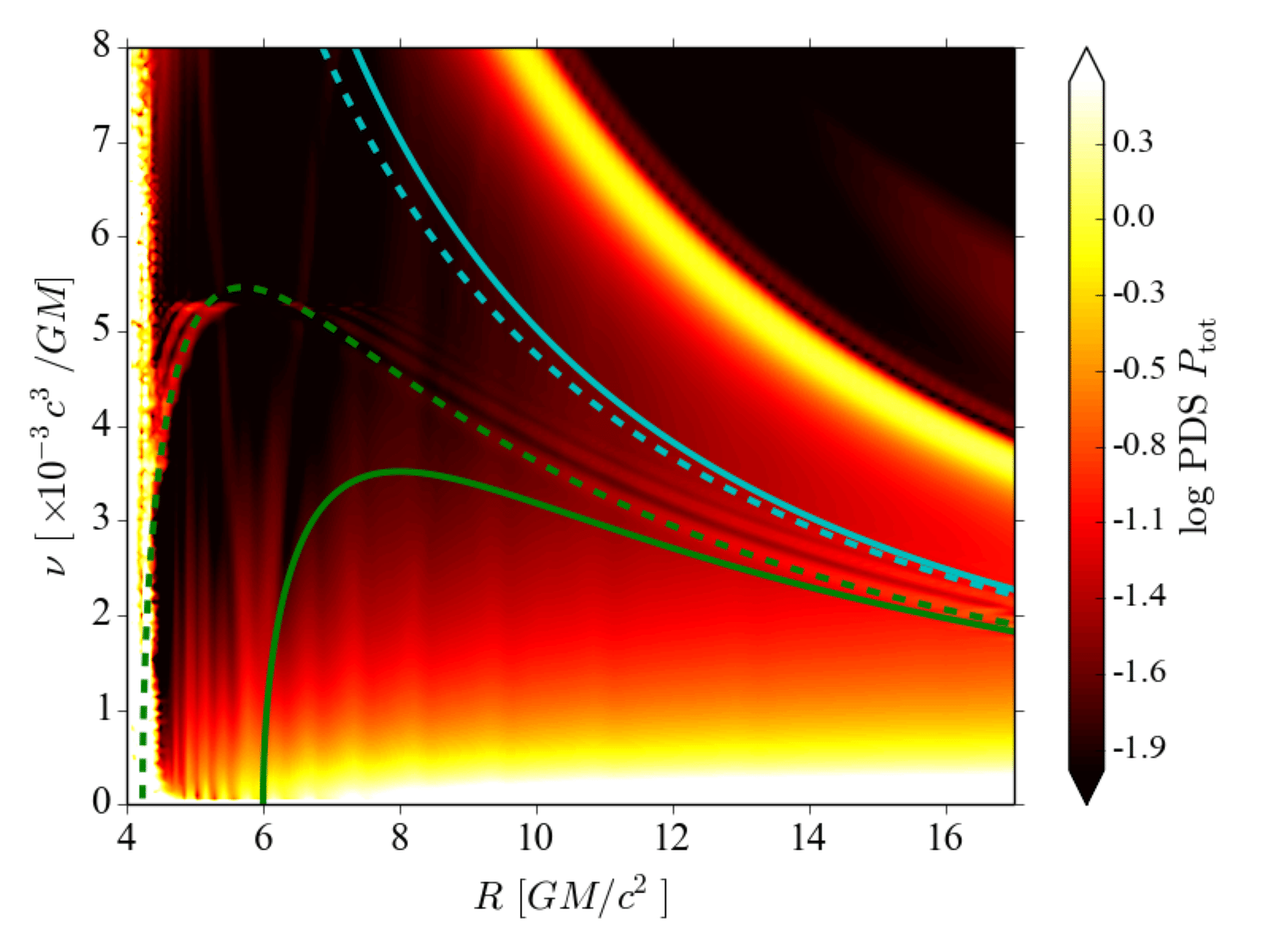}
\includegraphics[width=\columnwidth]{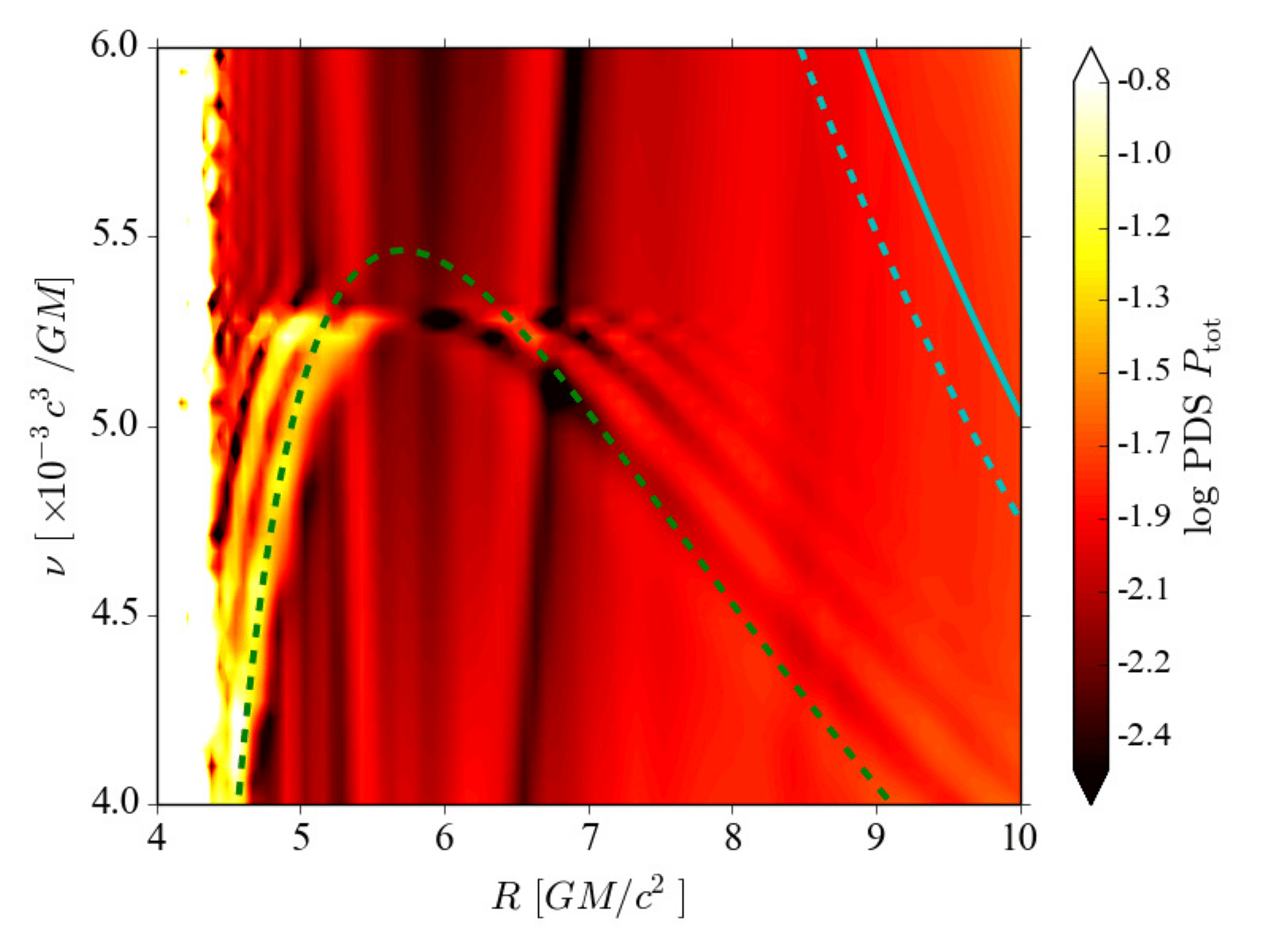}
\caption{Left: PDS of the total midplane pressure for simulation S01Ea5. Right: Zoomed in region of left panel.}
\label{fig:bhspin1}
\end{center}
\end{figure*}
In Fig.~\ref{fig:local_a0p5}, we present the {\it local} PDS of the mass accretion rate at $r = 5.6$, $5.7$, and $5.8\,r_\mathrm{g}$, which provides further evidence of a trapped ${\it g}$-mode and its interference with a ${\it p}$-mode. The maximum power in the magnified subplot is close to $r = 5.6\,r_\mathrm{g}$ (green dashed curve), which is slightly lower than the radius of the maximum radial epicyclic frequency. Similar to the non-rotating black hole case, the prominent power in the green, dashed curve gets enhanced by ${\it p}$-waves originating from the inner regions ($\nu \geq \tilde{\kappa}$) of the disk and penetrating the evanescent region ($\nu \le \tilde{\kappa}$). The power at $r = 5.46\,r_\mathrm{g}$ (radius of the maximum radial epicyclic frequency in this case), has a slightly lower magnitude compared to the one at $r = 5.6\,r_\mathrm{g}$. This again suggests the constructive coupling of ${\it p}$-waves and ${\it g}$-waves leading to larger power at $r = 5.6\,r_\mathrm{g}$. 
The power in peak at $r = 5.8\,r_\mathrm{g}$ is approximately $25\%$ smaller compared to the power at $r = 5.6$ and $5.7\,r_\mathrm{g}$. This indicates that there was no power leakage due to a ${\it p}$-mode originating at larger radii, $r>5.7$. To the contrary, the ${\it g}$-mode apparently excites a  ${\it p}-$wave at $r>6.8\,r_\mathrm{g}$. A similar feature, i.e., power decreasing from small to large radii, is seen in the right panel of Fig.~\ref{fig:bhspin}. This argues in favor of the presence of trapped ${\it g}$-mode (together with ${\it p}$-mode) oscillations also being present in the Schwarzschild S01E case, though not as easily distinguishable from oscillations occurring at the maximum radial epicyclic frequency. 

The total pressure PDS (Fig.~\ref{fig:bhspin1}) for the spinning black hole also shows evidence of diskoseismic oscillations, just as in the non-rotating black hole case. We note that the power appears to originate from the inner region with $\nu\ge \tilde{\kappa}(r)$, and then leak into the evanescent region with the same $\nu$ now satisfying $\nu\le\tilde{\kappa}(r)$. The ${\it g}$-and ${\it p}$-mode oscillations have clearly adapted to the correct radial epicyclic frequency (compare the solid and dashed curves in both Fig.~\ref{fig:bhspin} and \ref{fig:bhspin1}).

\begin{figure}
\begin{center}
\includegraphics[width=\columnwidth]{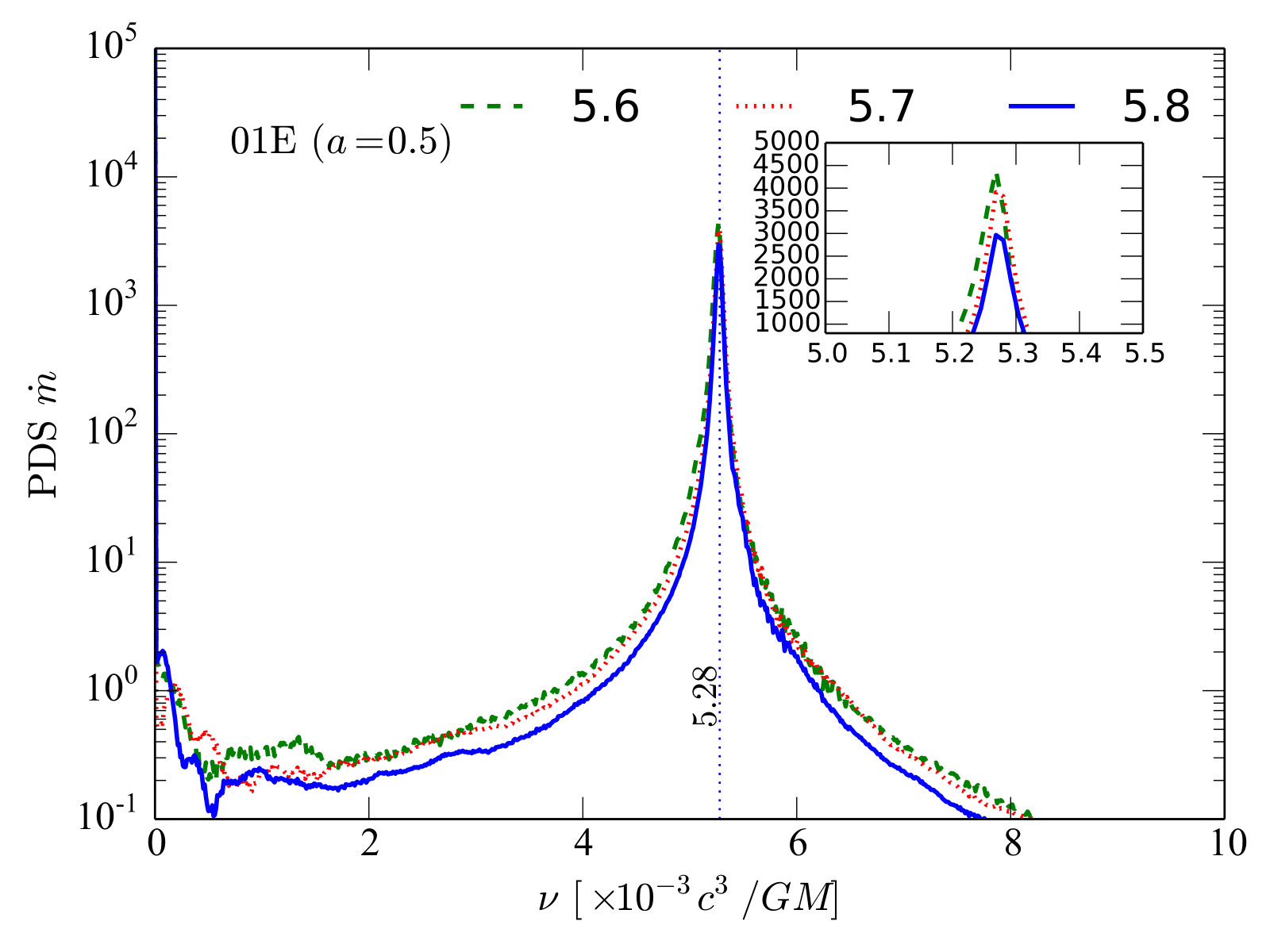}
\caption{PDS of local mass accretion rate for simulation S01Ea5 at $r = 5.6$, 5.7, and $5.8\,r_\mathrm{g}$.}
\label{fig:local_a0p5}
\end{center}
\end{figure}
\subsection{Breathing Oscillations}
\label{sec:breathing}

Accretion flows are prone to breathing oscillations as reported in \citet{Mishra17} based on studies of geometrically thick disks. For thin disks, the expressions for vertical oscillations, including the breathing oscillation, can be found in \cite{Bollimpalli17}, the breathing mode frequency for a $\gamma=5/3$ polytrope being $\nu=\sqrt{8/3}\,\nu_\perp$. In the numerical simulations of thin disks reported here, we find breathing oscillations at all radii larger than the ISCO. This is best seen in the left panels of Fig.~\ref{fig:bhspin} and Fig.~\ref{fig:bhspin1}, where we observe curves of large power, at frequencies $\nu(r) \approx 1.63\nu_\perp(r)$, starting from the outer regions of the disk all the way down to the inner edge. The vertical epicyclic frequency $\nu_\perp$ is shown by the dashed cyan curve. A detailed study of these breathing oscillations for the S01E case was reported in \citet{Mishra18}, who showed that the vertical motions in the disk at these frequencies match the predicted eigenfunctions. Thus, the identification of the breathing mode is unambiguous. One important conclusion of that paper was that the ratio of the breathing oscillation frequency to the vertical epicyclic frequency was approximately 5:3 (note that $\sqrt{8/3}\approx1.63$ is very close to $5/3\approx 1.67$). That conclusion also holds true for the spinning black hole simulation, S01Ea5, reported here.
\begin{figure*}
\begin{center}
\includegraphics[width=\columnwidth]{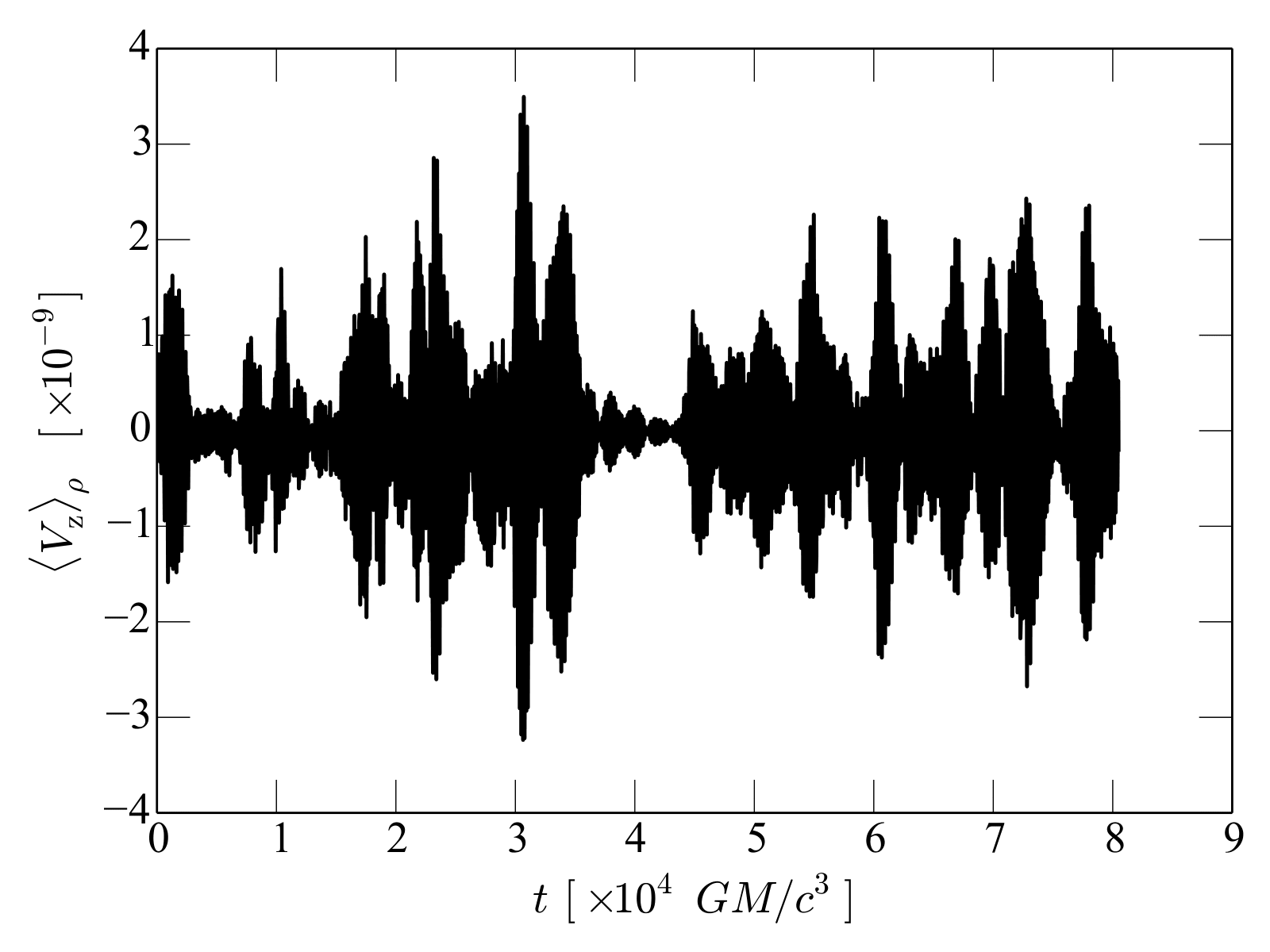}
\includegraphics[width=\columnwidth]{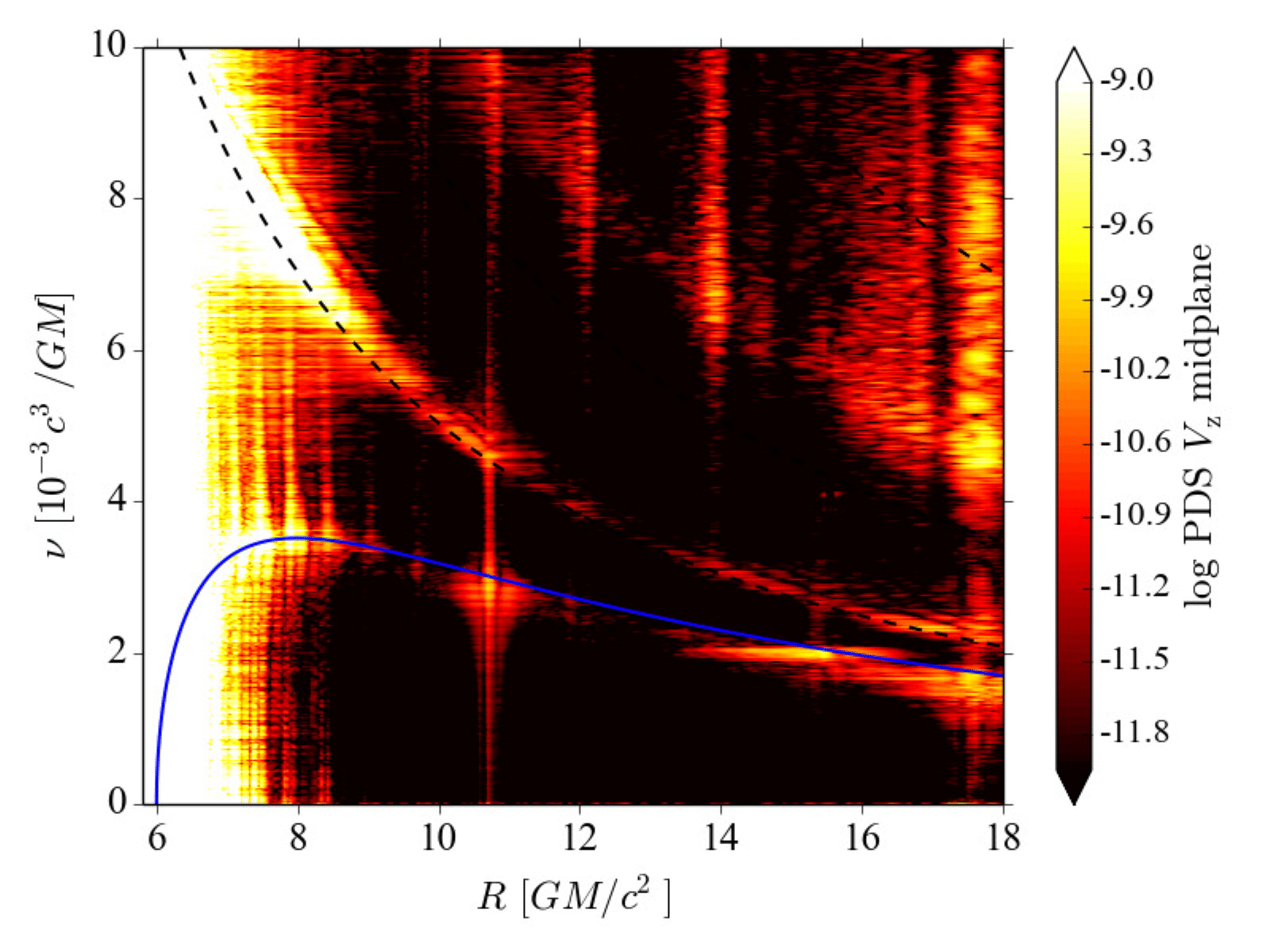}
\caption{Left: Time variability of the vertical velocity $\langle V^\mathrm{z}\rangle_\rho$ at $r = 8\,r_\mathrm{g} $ for simulation S01E. Right: PDS of the vertical midplane fluid velocity for a range of radii.}
\label{fig:time_var}
\end{center}
\end{figure*}

\begin{figure*}
\begin{center}
\includegraphics[width=\columnwidth]{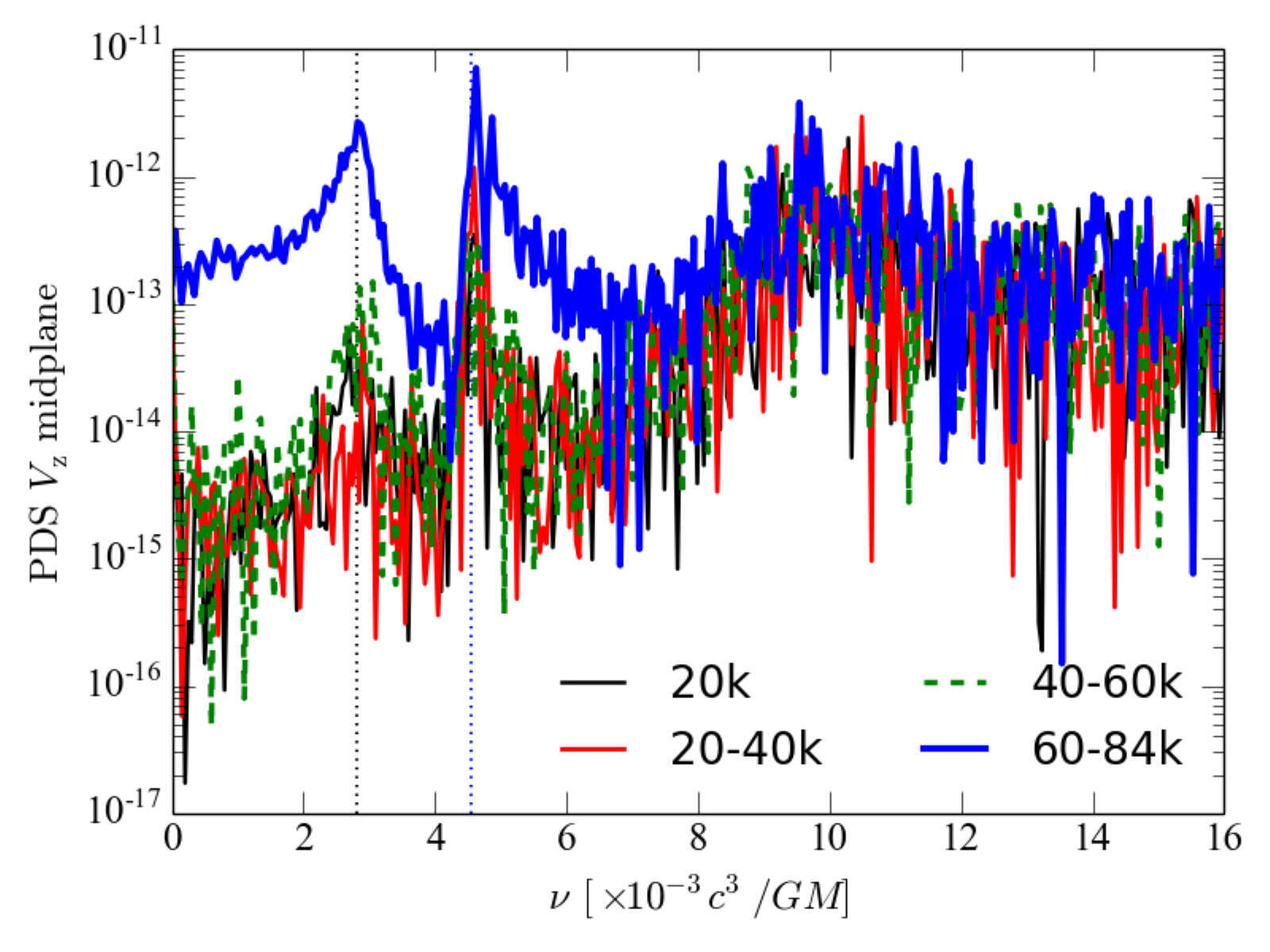}
\caption{PDS of the vertical velocity $\langle V^\mathrm{z}\rangle_\rho$ at $r = 10.8\,r_\mathrm{g}$ over various time windows. The time windows cover the first $2\times 10^4\,t_\mathrm{g}$ (solid black), $2-4\times 10^4\,t_\mathrm{g}$ (solid red), $4-6\times 10^4\,t_\mathrm{g}$ (dashed green) and $6-8.4\times 10^4\,t_\mathrm{g}$ (solid blue).}
\label{fig:local_spin}
\end{center}
\end{figure*}

\begin{figure*}
\begin{center}
\includegraphics[width=\columnwidth]{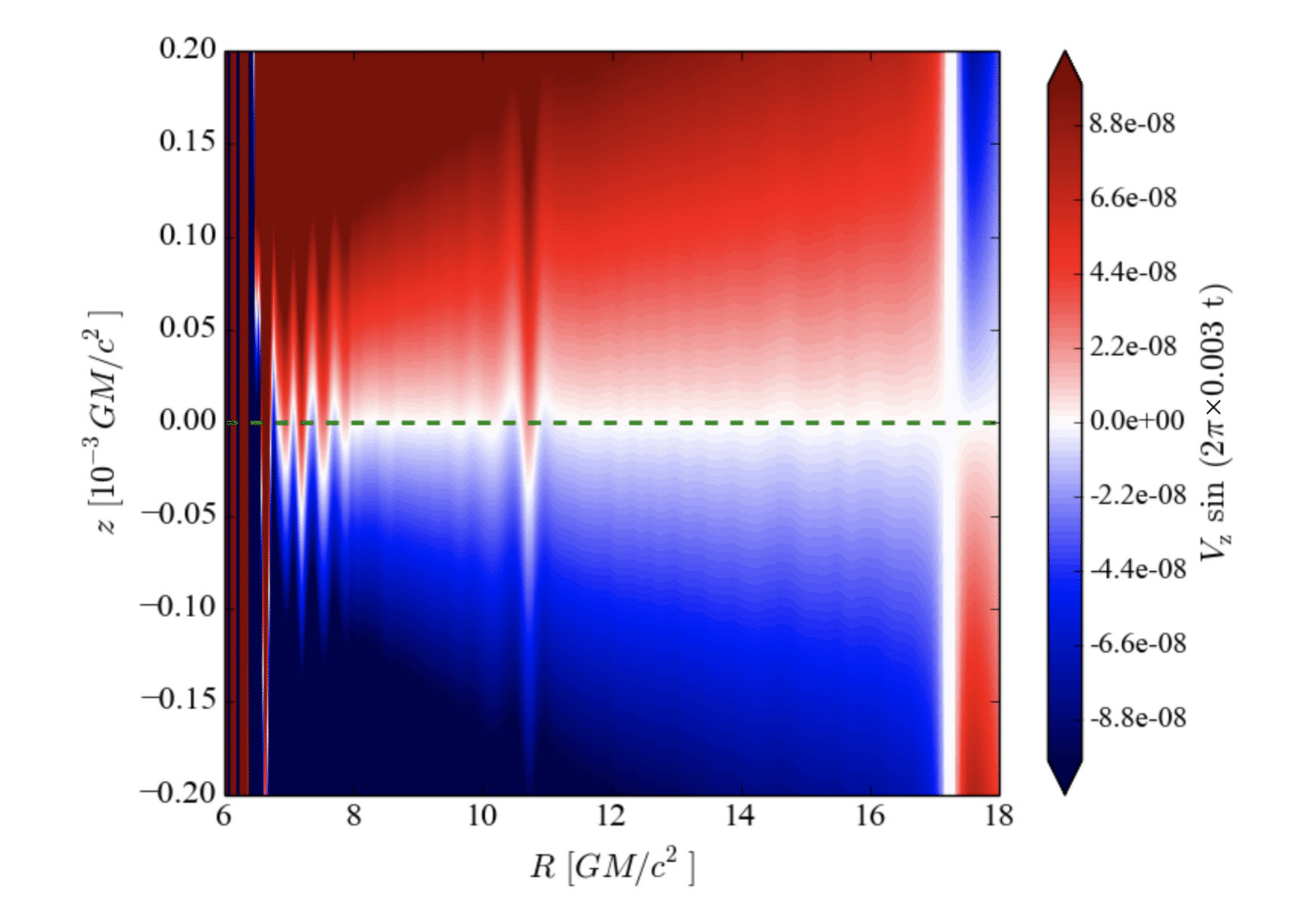}
\includegraphics[width=0.95\columnwidth]{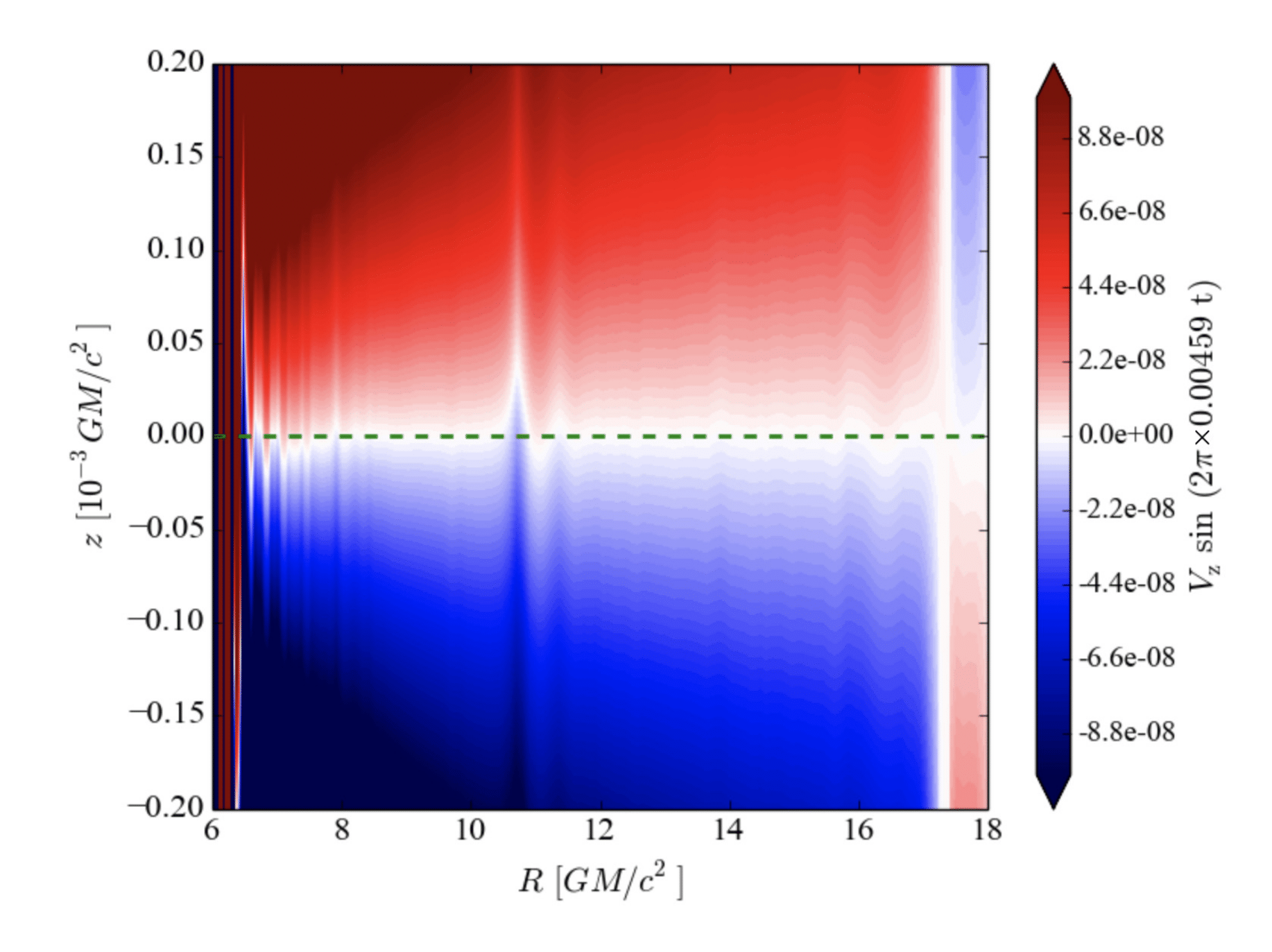}
\caption{Left: Eigenfunction of the vertical velocity evaluated at the frequency of $\nu = 3.00\times 10^{-3}\,t_\mathrm{g}^{-1}$. Right: Eigenfunction of the vertical velocity evaluated at the frequency of $\nu = 4.59\times 10^{-3}\,t_\mathrm{g}^{-1}$. The eigenfunctions are computed using the entire data set of simulation S01E.}
\label{fig:eigenfunc}
\end{center}
\end{figure*}
\subsection{Resonant oscillations at 3:2 frequency ratio for non-rotating black hole}
\label{sec:resonance}
 
A detailed study of the vertical component of the fluid three velocity shows one of the most striking oscillation features. In the left panel of Fig.~\ref{fig:time_var}, we show the time evolution of the density weighted vertical component of velocity $\langle V_\mathrm{z}\rangle_\rho$ at $r = 8\,r_\mathrm{g}$ for simulation S01E (similar results are seen for the spinning black hole case as well). It clearly shows fluid oscillations with small positive and negative velocities. This suggests obvious vertical oscillations across the disk midplane. The presence of the wave trains suggest additional oscillation features, which can best be analyzed by looking at the Fourier transform in the right panel. We find evidence of vertical oscillations from the bright strip of power tracing the black dashed curve of Keplerian orbital frequency, equal in this Schwarzschild case to the vertical epicyclic frequency, $\nu_{\perp}=\tilde{\Omega}(r)$.  Additionally, we see power tracing the radial epicyclic frequency, $\tilde{\kappa}(r)$ (solid blue curve). Examining more closely, we see enhanced power in the vertical oscillations relative to the oscillations occurring close to radial epicyclic frequency at small radii, but the opposite at larger radii. There is very little power along the radial epicyclic curve in the interval  $9\,r_\mathrm{g}<r<14 r_\mathrm{g}$, this makes the feature at $r = 10.8\,r_\mathrm{g}$, with excess power at both the radial and the vertical epicyclic frequency, quite striking.

The resonance models put forward to explain the observed $3:2$ frequency ratio in twin-peak QPOs \citep{Abramowicz01} predict that resonance will occur at certain specific radii; for example, $r = 10.8\,r_\mathrm{g}$ gives a $3:2$ frequency ratio between the vertical and radial epicyclic frequencies for a Schwarzschild black hole, and this has been exploited to predict resonance between the vertical and radial eigenmodes of oscillating slender tori \citep{Bursa04,Kluzniak05}. Whether one expects similar motions in a thin disk is unclear,  but given that both the radial and vertical epicyclic frequencies are associated with excess power over a wide range of radii, as seen in the right panel of  Fig.~\ref{fig:time_var}, one could expect that the corresponding motions could similarly be coupled in a non-linear resonance. This seems to be indeed the case, with the appearance of the feature  at $r = 10.8\,r_\mathrm{g}$. We plot PDSs at this specific radius for different time intervals in Fig.~\ref{fig:local_spin}. The largest power is in the time window $t = 60000$ to $80000\,r_\mathrm{g}$, although, the power grew steadily during the earlier evolution (see the solid black, solid red and dashed green curves).

In the right panel of Fig. 10, at larger radii, we see horizontal strips of power at fixed frequencies. The most prominent of these is the clearly apparent strip of power at $14 r_\mathrm{g} < r < 16 r_\mathrm{g}$ below the radial epicyclic frequency, with no obvious counterpart at a higher frequency. Apparently, in this radial domain, a wide annular fragment of the disk is executing coherent motion with a nonzero midplane vertical velocity oscillation, at a frequency corresponding to the radial epicyclic frequency at $r=16 r_\mathrm{g}$. The nature of this oscillation remains unknown. At larger radii still, the annulus at $17 r_\mathrm{g} < r < 18 r_\mathrm{g}$ seems to be executing vertical motions at more than one frequency.

The eigenfunctions of $\langle V_z \rangle_\rho$ in Fig.~\ref{fig:eigenfunc} quantitatively illustrate the fluid motion within the disk for individual oscillations. In this approach we pick the frequency of a specific oscillation and compute the time integral of a sinusoid multiplied by the fluid variable in question. Specifically, Fig.~\ref{fig:eigenfunc} shows the time-integrated vertical velocity in the $(r,z)$ plane evaluated at the frequency of $\nu = 3.00\times 10^{-3}\,t_\mathrm{g}^{-1}$ i.e., the radial epicyclic frequency at $r= 10.8\,r_\mathrm{g}$ (left panel) and the corresponding vertical epicyclic frequency of $\nu = 4.59\times 10^{-3}\,t_\mathrm{g}^{-1}$ (right panel) over the entire duration of simulation S01E. Blue color regions at the $z = 0$ plane correspond to downwards fluid motion, i.e., in the negative $z$ direction, whereas red color corresponds to upwards motion. We clearly note vertical motion across the midplane around $r = 10.8\,r_\mathrm{g}$ at both the radial and vertical epicyclic frequencies. Remember that, in addition to seeing the vertical and radial oscillations occurring with a 3:2 frequency ratio, we also capture breathing oscillations occurring at all the radii at approximately 5:3 frequency ratio with the vertical epicyclic frequency \citep{Mishra18}. The frequencies we chose for generating the eigenfunctions in Fig.12 capture these breathing oscillations at large radii ($r \approx 18\,r_\mathrm{g}$, right edge of figure).
\begin{figure*}
\begin{center}
\includegraphics[width=\columnwidth]{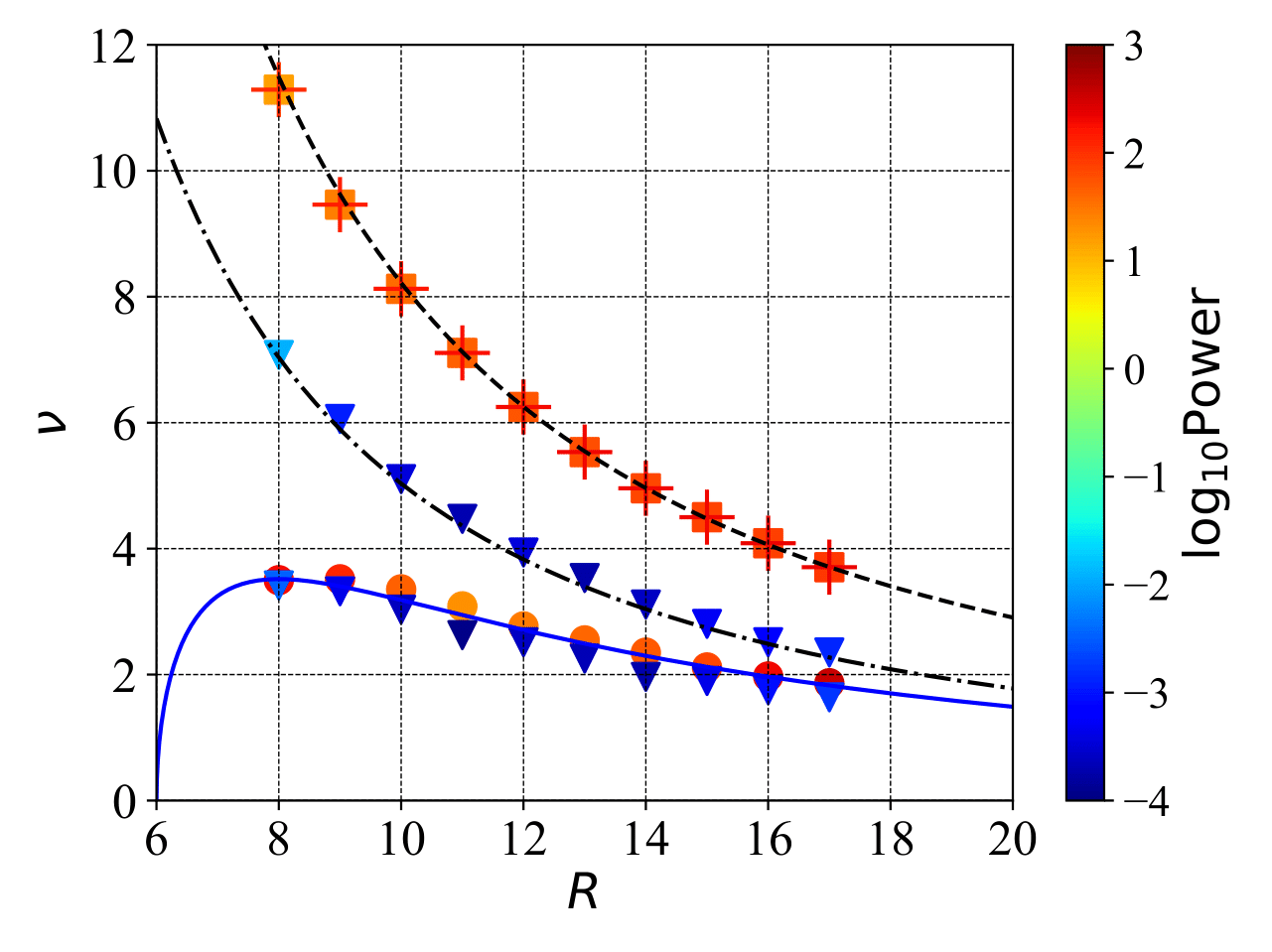}
\caption{Comparison of the full set of modes reported in this work. Triangles show oscillations seen in PDS of the vertical midplane velocity, filled circles show oscillations seen in the PDS of the radial midplane velocity, and filled squares and `plus' symbols show oscillations seen in PDS of the radial midplane velocity and the vertical velocity integrated above the midplane, respectively (to show the breathing oscillations). The color bar shows power normalized to the initial radial velocity amplitude. 
The solid, dot-dashed, and dashed curves are plots of the radial epicyclic, vertical epicyclic, and the breathing oscillation frequency, respectively.}
\label{fig:dsfig16}
\end{center}
\end{figure*}
\section{Comparison with previous studies}
\label{sec:comparison}

In a series of papers \citet{Reynolds09,ONeill09} carried out the first detailed study of diskoseismic oscillations in hydrodynamic and MHD simulations of geometrically thin disks. Our study differs in several important ways, including the initial accretion disk profiles, our full treatment of the general theory of relativity, and self-consistent radiative transfer (with a frequency-integrated approximation). Despite these differences, we reproduce all of the findings of the hydrodynamical simulations in \citet{Reynolds09,ONeill09}, while also discovering modes of oscillation that were not reported in their study.
Specifically, \citet{Reynolds09,ONeill09} found evidence for ${\it g}$-modes (in the Paczynski--Wiita pseudo-potential) in their viscous hydrodynamic simulations, but not in the MHD simulations, as well as ${\it p}$-modes. Our hydrodynamic simulations, performed in GR, correspond to their viscous case, and we confirm the presence of ${\it g}$-modes, both in the Schwarzschild- and Kerr-metric simulations. The frequency of this mode clearly corresponds to the analytic predictions of the linear theory \citep{Perez97}. ${\it p}$-modes are also abundantly present.

Given the fact that our S1E, S3E, and S10E simulations collapse into much cooler gas-pressure-dominated disks \citep[see][]{Fragile18}, and that the S01E disk is in vertical hydrostatic equilibrium with an initial Gaussian density profile, our $\dot{m}(r)$ plots should correspond to the midplane radial velocity plots of \citet{ONeill09}. Indeed, both studies reveal white or red noise at $r<6\,r_\mathrm{g}$, i.e., below the ISCO, as well as a strip of power at a frequency close to $\tilde{\kappa}_\mathrm{max}$ (seen more clearly in the left panels of Figs.~\ref{fig:dsfig3} and~\ref{fig:dsfig4}). However, the latter characteristic is only present at high values of the viscosity parameter, $\alpha>0.07$ in the \citet{ONeill09} study (see their Fig.~6), while in our simulations it is present even at $\alpha=0.02$. It could be that a combination of general relativistic effects and radiation enhance certain oscillating features. There is also moderately higher power for frequencies equal to and slightly larger than $\tilde{\kappa}(r)$ at larger radii \citep[$r>16\,r_\mathrm{g}$ in][and at all radii in our simulations]{ONeill09}.
 A number of  features in our global simulations had not  been reported in previous numerical studies on diskoseismic modes \citep{Reynolds09,ONeill09}. In addition to the horizontal strip of power at the frequency of about $\tilde{\kappa}_\mathrm{max}$, we can clearly see harmonics at $2\tilde{\kappa}_\mathrm{max}$ and $3\tilde{\kappa}_\mathrm{max}$, as well as the corresponding harmonics of the excess power at $\tilde{\kappa}(r)$ for $r<8\,r_\mathrm{g}$.

 We also confirm the presence of ${\it p}$-modes at radii larger than that of the maximum of the radial epicyclic frequency.  With our superior resolution,  to the right of the radial epicyclic curve we clearly see standing ${\it p}$-waves with $\nu<\tilde{\kappa}_\mathrm{max}$. The previously reported excess power close to the inner edge of the disk \citep{ONeill09}, is clearly present in our simulations. Some of this power we interpret as trapped ${\it p}$-modes.

We identify the breathing mode, and here our interpretation differs from \citet{Reynolds09}, who discuss a similar feature in terms of high frequency, purely vertical acoustic waves in a locally vertically isothermal atmosphere.
   Our atmosphere is not vertically isothermal, so we do not expect these specific acoustic modes to be applicable.   
 Also the spectrum of these oscillations is different from that of the vertical oscillations to which the breathing mode belongs.
We rewrite Eq. 4 of \citet{Reynolds09} in the form
\begin{equation}
\label{eq:chris09}
\nu^2 = \frac{(m+1)n + m}{n}\nu^2_\perp ,
\end{equation}
and compare it with the vertical acoustic-inertial oscillations discussed in \citet{Perez97, Bollimpalli17, Bollimpalli19},
\begin{equation}
\label{eq:bm}
\nu^2 = \frac{m(m+2n-1)}{2n}\nu^2_\perp ,
\end{equation} 
where the polytropic index is $\gamma = 1 + {1}/{n}$, and $m=0, 1, 2, ...$ is the mode number.
Specializing to a $\gamma = 5/3$ polytrope ($n=3/2$), we can rewrite Eq.~\ref{eq:chris09} as $\nu^2/\nu^2_\perp =1+(5/3)m=1, 8/3, 13/3, 6,...$, while  Eq.~\ref{eq:bm} yields $\nu^2/\nu^2_\perp =m(m +2)/3 = 0,1,8/3,5,8,...$ for $m=0,1,2,3,4,...$
Clearly, while the first two non-zero eigenfrequencies happen to coincide in the series, these two classes of oscillations are completely different. This suggests that $m=2$ vertical ${\it p}-$modes reported in \citet{Reynolds09} do not belong to the same class of modes that we reported in \citet{Mishra18}. As stated above, the identification of the power at $1.63\,\nu_\perp$ as a manifestation of breathing oscillations is unambiguous following an analysis of the wavefunctions \citep{Mishra18}.
Further details on classes of oscillation modes can be found in Section 4.2.1 in \citet{Kato16}.
\section{Conclusions}
\label{sec:conclusion}

We have analyzed the high-frequency timing variability of simulated radiation- and gas-pressure-dominated thin accretion disks. Using the approximation of an $\alpha$-viscosity prescription instead of directly simulating magnetically driven turbulence, we see a rich set of oscillation features. 

\begin{itemize}
\item[1.] A coherent oscillation occurring close to the maximum radial epicyclic frequency, $\tilde{\kappa}_\mathrm{max}=3.5\times 10^{-3}\, t_\mathrm{g}^{-1}$ ($=108\,$Hz for a $6.6 M_\odot$ non-rotating black hole) is present in all of our Schwarzschild simulations. This is consistent with the presence of a ${\it g}$-mode. In the radiation-pressure-dominated simulations, which are subject to thermal collapse, such as S1E, there is also a ${\it p}$-wave oscillation at a frequency just above $\tilde{\kappa}_\mathrm{max}$ throughout the disk (at all radii). 

\item[2.] In the Kerr-metric simulation (S01Ea5), there is a great deal of power in the inner disk at a frequency of $\nu\approx 5.28\times 10^{-3}\,t_\mathrm{g}^{-1}$, well below the maximum radial epicyclic frequency value of  $5.46\times 10^{-3}\,t_\mathrm{g}^{-1}$. Both, the radial distribution of power and the frequency are consistent with this being a ${\it g}$-mode, as predicted by \citet{Perez97}.

\item[3.] The stable, gas-pressure-dominated simulations, S01E and S01Ea5, show much quieter PDS. There is no detectable high-frequency power in the pressure below the radial epicyclic frequency curve (cf. Fig.~\ref{fig:dsfig6}), consistent with the predictions of the linear theory of diskoseismology \citep{Wagoner99,Kato2001}. In the mass flux PDS, there is strong evidence for standing ${\it p}$-mode waves outside the radial epicyclic frequency curve, always with angular frequencies below $\tilde{\kappa}_\mathrm{max}$, with a rather sharp power cut-off at $\nu=\tilde{\kappa}_\mathrm{max}$. 

\item[4.] Oscillations occurring at the breathing oscillation frequency  $\approx1.63\,\nu_\perp$ are seen in all of our reported simulations.

\item[5.] Evidence of pairs of oscillations occurring at a 3:2 frequency ratio is seen in all of our simulations. For the first time in a simulation we see evidence of a 3:2 resonance between the radial and vertical epicyclic frequencies, occurring at $r=10.8 r_\mathrm{g}$ in the Schwarzschild case.

\end{itemize}
A complete picture of all the reported disk oscillations in this study is provided in Fig.~\ref{fig:dsfig16}.  Although multiple modes are identified, the power is mainly concentrated in the breathing  and radial epicyclic oscillations. The vertical oscillations occurring near the vertical and radial epicyclic frequencies have six orders of magnitude less power. 
A comparison of the radial distribution of power shows an increase of power with radius for the breathing mode, while the power in the vertical and radial epicyclic modes generally decreases with radius.

\section*{Acknowledgements}
BM thanks to Omer Blaes for discussions and suggestions. BM acknowledges support from NASA Astrophysics Theory Program grants  NNX16AI40G and NNX17AK55G, and from NSF grant AST-1411879. It used resources from the Extreme Science and Engineering Discovery Environment (XSEDE), which is supported by National Science Foundation grant number ACI-1053575 and PROMETHEUS supercomputer in the PL-Grid infrastructure in Poland. BM and WK acknowledge support from Polish NCN grants 2013/08/A/ST9/00795 and 2014/15/N/ST9/04633.
\bibliographystyle{mnras}
\bibliography{refs}

\begin{thebibliography}{}
\makeatletter
\relax
\def\mn@urlcharsother{\let\do\@makeother \do\$\do\&\do\#\do\^\do\_\do\%\do\~}
\def\mn@doi{\begingroup\mn@urlcharsother \@ifnextchar [ {\mn@doi@}
  {\mn@doi@[]}}
\def\mn@doi@[#1]#2{\def\@tempa{#1}\ifx\@tempa\@empty \href
  {http://dx.doi.org/#2} {doi:#2}\else \href {http://dx.doi.org/#2} {#1}\fi
  \endgroup}
\def\mn@eprint#1#2{\mn@eprint@#1:#2::\@nil}
\def\mn@eprint@arXiv#1{\href {http://arxiv.org/abs/#1} {{\tt arXiv:#1}}}
\def\mn@eprint@dblp#1{\href {http://dblp.uni-trier.de/rec/bibtex/#1.xml}
  {dblp:#1}}
\def\mn@eprint@#1:#2:#3:#4\@nil{\def\@tempa {#1}\def\@tempb {#2}\def\@tempc
  {#3}\ifx \@tempc \@empty \let \@tempc \@tempb \let \@tempb \@tempa \fi \ifx
  \@tempb \@empty \def\@tempb {arXiv}\fi \@ifundefined
  {mn@eprint@\@tempb}{\@tempb:\@tempc}{\expandafter \expandafter \csname
  mn@eprint@\@tempb\endcsname \expandafter{\@tempc}}}

\bibitem[\protect\citeauthoryear{{Abramowicz} \& {Klu{\'z}niak}}{{Abramowicz}
  \& {Klu{\'z}niak}}{2001}]{Abramowicz01}
{Abramowicz} M.~A.,  {Klu{\'z}niak} W.,  2001, \mn@doi [\aap]
  {10.1051/0004-6361:20010791}, \href
  {http://adsabs.harvard.edu/abs/2001A%26A...374L..19A} {374, L19}

\bibitem[\protect\citeauthoryear{{Bollimpalli} \& {Klu{\'z}niak}}{{Bollimpalli}
  \& {Klu{\'z}niak}}{2017}]{Bollimpalli17}
{Bollimpalli} D.~A.,  {Klu{\'z}niak} W.,  2017, \mn@doi [\mnras]
  {10.1093/mnras/stx2140}, \href
  {https://ui.adsabs.harvard.edu/abs/2017MNRAS.472.3298B} {472, 3298}

\bibitem[\protect\citeauthoryear{{Bollimpalli}, {Wielgus}, {Abarca}  \&
  {Klu{\'z}niak}}{{Bollimpalli} et~al.}{2019}]{Bollimpalli19}
{Bollimpalli} D.~A.,  {Wielgus} M.,  {Abarca} D.,   {Klu{\'z}niak} W.,  2019,
  \mn@doi [\mnras] {10.1093/mnras/stz1597}, \href
  {https://ui.adsabs.harvard.edu/abs/2019MNRAS.tmp.1529B} {}

\bibitem[\protect\citeauthoryear{{Bursa}, {Abramowicz}, {Karas}  \&
  {Klu{\'z}niak}}{{Bursa} et~al.}{2004}]{Bursa04}
{Bursa} M.,  {Abramowicz} M.~A.,  {Karas} V.,   {Klu{\'z}niak} W.,  2004,
  \mn@doi [\apjl] {10.1086/427167}, \href
  {http://adsabs.harvard.edu/abs/2004ApJ...617L..45B} {617, L45}

\bibitem[\protect\citeauthoryear{{Chen} \& {Taam}}{{Chen} \&
  {Taam}}{1995}]{Chen95}
{Chen} X.,  {Taam} R.~E.,  1995, \mn@doi [\apj] {10.1086/175360}, \href
  {http://adsabs.harvard.edu/abs/1995ApJ...441..354C} {441, 354}

\bibitem[\protect\citeauthoryear{{Fragile}, {Etheridge}, {Anninos}, {Mishra}
  \& {Klu{\'z}niak}}{{Fragile} et~al.}{2018}]{Fragile18}
{Fragile} P.~C.,  {Etheridge} S.~M.,  {Anninos} P.,  {Mishra} B.,
  {Klu{\'z}niak} W.,  2018, \mn@doi [\apj] {10.3847/1538-4357/aab788}, \href
  {http://adsabs.harvard.edu/abs/2018ApJ...857....1F} {857, 1}

\bibitem[\protect\citeauthoryear{{Giussani}, {Klu{\'z}niak}  \&
  {Mishra}}{{Giussani} et~al.}{2014}]{Giussani14}
{Giussani} L.,  {Klu{\'z}niak} W.,   {Mishra} B.,  2014, in {Ztuchl{\'i}k} Z.,
  {T{\"o}r{\"o}k} G.,   {Pech{\'a}\v{c}ek} T.,  eds, Proceedings of RAGtime
  14-16: Workshop on blackholes and neutron stars, arXiv:1503.05546. pp 93--98
  (\mn@eprint {arXiv} {1503.05546})

\bibitem[\protect\citeauthoryear{{Homan}, {Miller}, {Wijnands}, {van der Klis},
  {Belloni}, {Steeghs}  \& {Lewin}}{{Homan} et~al.}{2005}]{Homan05}
{Homan} J.,  {Miller} J.~M.,  {Wijnands} R.,  {van der Klis} M.,  {Belloni} T.,
   {Steeghs} D.,   {Lewin} W.~H.~G.,  2005, \mn@doi [\apj] {10.1086/424994},
  \href {http://adsabs.harvard.edu/abs/2005ApJ...623..383H} {623, 383}

\bibitem[\protect\citeauthoryear{{Honma}, {Matsumoto}  \& {Kato}}{{Honma}
  et~al.}{1992}]{Honma92}
{Honma} F.,  {Matsumoto} R.,   {Kato} S.,  1992, \pasj, \href
  {http://adsabs.harvard.edu/abs/1992PASJ...44..529H} {44, 529}

\bibitem[\protect\citeauthoryear{{Kato}}{{Kato}}{2001}]{Kato2001}
{Kato} S.,  2001, \mn@doi [\pasj] {10.1093/pasj/53.1.1}, \href
  {http://adsabs.harvard.edu/abs/2001PASJ...53....1K} {53, 1}

\bibitem[\protect\citeauthoryear{{Kato}}{{Kato}}{2016}]{Kato16}
{Kato} S.,  ed. 2016, {Oscillations of Disks}  Astrophysics and Space Science
  Library Vol. 437, \mn@doi{10.1007/978-4-431-56208-5.
}

\bibitem[\protect\citeauthoryear{{Kato} \& {Fukue}}{{Kato} \&
  {Fukue}}{1980}]{Kato80}
{Kato} S.,  {Fukue} J.,  1980, \pasj, \href
  {https://ui.adsabs.harvard.edu/abs/1980PASJ...32..377K} {32, 377}

\bibitem[\protect\citeauthoryear{{Kato}, {Fukue}  \& {Mineshige}}{{Kato}
  et~al.}{2008}]{Kato08}
{Kato} S.,  {Fukue} J.,   {Mineshige} S.,  2008, {Black-Hole Accretion Disks
  --- Towards a New Paradigm ---}

\bibitem[\protect\citeauthoryear{{Klu{\'z}niak}}{{Klu{\'z}niak}}{2005}]{Kluzniak05}
{Klu{\'z}niak} W.,  2005, \mn@doi [Astronomische Nachrichten]
  {10.1002/asna.200510420}, \href
  {http://adsabs.harvard.edu/abs/2005AN....326..820K} {326, 820}

\bibitem[\protect\citeauthoryear{{Klu\'zniak} \& {Abramowicz}}{{Klu\'zniak} \&
  {Abramowicz}}{2001}]{Kluzniak01}
{Klu\'zniak} W.,  {Abramowicz} M.~A.,  2001, Acta Physica Polonica B, \href
  {http://adsabs.harvard.edu/abs/2001AcPPB..32.3605K} {32, 3605}

\bibitem[\protect\citeauthoryear{{Kluzniak} \& {Abramowicz}}{{Kluzniak} \&
  {Abramowicz}}{2002}]{Kluzniak02}
{Kluzniak} W.,  {Abramowicz} M.~A.,  2002, ArXiv Astrophysics e-prints, \href
  {http://adsabs.harvard.edu/abs/2002astro.ph..3314K} {}

\bibitem[\protect\citeauthoryear{{Klu{\'z}niak} \& {Lee}}{{Klu{\'z}niak} \&
  {Lee}}{2002}]{KL2002}
{Klu{\'z}niak} W.,  {Lee} W.~H.,  2002, \mn@doi [\mnras]
  {10.1046/j.1365-8711.2002.05819.x}, \href
  {http://adsabs.harvard.edu/abs/2002MNRAS.335L..29K} {335, L29}

\bibitem[\protect\citeauthoryear{{Klu{\'z}niak}, {Abramowicz}  \&
  {Lee}}{{Klu{\'z}niak} et~al.}{2004}]{Kluzniak04}
{Klu{\'z}niak} W.,  {Abramowicz} M.~A.,   {Lee} W.~H.,  2004, in {Kaaret} P.,
  {Lamb} F.~K.,   {Swank} J.~H.,  eds,  American Institute of Physics
  Conference Series Vol. 714, X-ray Timing 2003: Rossi and Beyond. pp 379--382
  (\mn@eprint {} {astro-ph/0402013}), \mn@doi{10.1063/1.1781058}

\bibitem[\protect\citeauthoryear{{Mihalas} \& {Mihalas}}{{Mihalas} \&
  {Mihalas}}{1984}]{Mihalas84}
{Mihalas} D.,  {Mihalas} B.~W.,  1984, {Foundations of radiation hydrodynamics}

\bibitem[\protect\citeauthoryear{{Milsom} \& {Taam}}{{Milsom} \&
  {Taam}}{1996}]{Milsom96}
{Milsom} J.~A.,  {Taam} R.~E.,  1996, \mn@doi [\mnras]
  {10.1093/mnras/283.3.919}, \href
  {http://adsabs.harvard.edu/abs/1996MNRAS.283..919M} {283, 919}

\bibitem[\protect\citeauthoryear{{Milsom} \& {Taam}}{{Milsom} \&
  {Taam}}{1997}]{Milsom97}
{Milsom} J.~A.,  {Taam} R.~E.,  1997, \mn@doi [\mnras]
  {10.1093/mnras/286.2.358}, \href
  {http://adsabs.harvard.edu/abs/1997MNRAS.286..358M} {286, 358}

\bibitem[\protect\citeauthoryear{{Mishra}, {Vincent}, {Manousakis}, {Fragile},
  {Paumard}  \& {Klu{\'z}niak}}{{Mishra} et~al.}{2017}]{Mishra17}
{Mishra} B.,  {Vincent} F.~H.,  {Manousakis} A.,  {Fragile} P.~C.,  {Paumard}
  T.,   {Klu{\'z}niak} W.,  2017, \mn@doi [\mnras] {10.1093/mnras/stx299},
  \href {http://adsabs.harvard.edu/abs/2017MNRAS.467.4036M} {467, 4036}

\bibitem[\protect\citeauthoryear{{Mishra}, {Klu{\'z}niak}  \&
  {Fragile}}{{Mishra} et~al.}{2019}]{Mishra18}
{Mishra} B.,  {Klu{\'z}niak} W.,   {Fragile} P.~C.,  2019, \mn@doi [\mnras]
  {10.1093/mnras/sty3124}, \href
  {https://ui.adsabs.harvard.edu/abs/2019MNRAS.483.4811M} {483, 4811}

\bibitem[\protect\citeauthoryear{{Morgan}, {Remillard}  \& {Greiner}}{{Morgan}
  et~al.}{1997}]{Morgan97}
{Morgan} E.~H.,  {Remillard} R.~A.,   {Greiner} J.,  1997, \mn@doi [\apj]
  {10.1086/304191}, \href {http://adsabs.harvard.edu/abs/1997ApJ...482..993M}
  {482, 993}

\bibitem[\protect\citeauthoryear{{Nowak} \& {Wagoner}}{{Nowak} \&
  {Wagoner}}{1991}]{Nowak91}
{Nowak} M.~A.,  {Wagoner} R.~V.,  1991, \mn@doi [\apj] {10.1086/170465}, \href
  {http://adsabs.harvard.edu/abs/1991ApJ...378..656N} {378, 656}

\bibitem[\protect\citeauthoryear{{Nowak} \& {Wagoner}}{{Nowak} \&
  {Wagoner}}{1992}]{Nowak92}
{Nowak} M.~A.,  {Wagoner} R.~V.,  1992, \mn@doi [\apj] {10.1086/171538}, \href
  {http://adsabs.harvard.edu/abs/1992ApJ...393..697N} {393, 697}

\bibitem[\protect\citeauthoryear{{Nowak} \& {Wagoner}}{{Nowak} \&
  {Wagoner}}{1993}]{Nowak93}
{Nowak} M.~A.,  {Wagoner} R.~V.,  1993, \mn@doi [\apj] {10.1086/173381}, \href
  {http://adsabs.harvard.edu/abs/1993ApJ...418..187N} {418, 187}

\bibitem[\protect\citeauthoryear{{O'Neill}, {Reynolds}  \& {Miller}}{{O'Neill}
  et~al.}{2009}]{ONeill09}
{O'Neill} S.~M.,  {Reynolds} C.~S.,   {Miller} M.~C.,  2009, \mn@doi [\apj]
  {10.1088/0004-637X/693/2/1100}, \href
  {http://adsabs.harvard.edu/abs/2009ApJ...693.1100O} {693, 1100}

\bibitem[\protect\citeauthoryear{{Okazaki}, {Kato}  \& {Fukue}}{{Okazaki}
  et~al.}{1987}]{Okazaki87}
{Okazaki} A.~T.,  {Kato} S.,   {Fukue} J.,  1987, \pasj, \href
  {http://adsabs.harvard.edu/abs/1987PASJ...39..457O} {39, 457}

\bibitem[\protect\citeauthoryear{{Ortega-Rodr{\'{\i}}guez}, {Silbergleit}  \&
  {Wagoner}}{{Ortega-Rodr{\'{\i}}guez} et~al.}{2002}]{Ortega02}
{Ortega-Rodr{\'{\i}}guez} M.,  {Silbergleit} A.~S.,   {Wagoner} R.~V.,  2002,
  \mn@doi [\apj] {10.1086/338685}, \href
  {https://ui.adsabs.harvard.edu/abs/2002ApJ...567.1043O} {567, 1043}

\bibitem[\protect\citeauthoryear{{Perez}, {Silbergleit}, {Wagoner}  \&
  {Lehr}}{{Perez} et~al.}{1997}]{Perez97}
{Perez} C.~A.,  {Silbergleit} A.~S.,  {Wagoner} R.~V.,   {Lehr} D.~E.,  1997,
  \mn@doi [\apj] {10.1086/303658}, \href
  {http://adsabs.harvard.edu/abs/1997ApJ...476..589P} {476, 589}

\bibitem[\protect\citeauthoryear{{Remillard}}{{Remillard}}{2004}]{Remillard04}
{Remillard} R.~A.,  2004, in {Kaaret} P.,  {Lamb} F.~K.,   {Swank} J.~H.,  eds,
   American Institute of Physics Conference Series Vol. 714, X-ray Timing 2003:
  Rossi and Beyond. pp 13--20, \mn@doi{10.1063/1.1780992}

\bibitem[\protect\citeauthoryear{{Remillard} \& {McClintock}}{{Remillard} \&
  {McClintock}}{2006}]{Remillard06}
{Remillard} R.~A.,  {McClintock} J.~E.,  2006, \mn@doi [\araa]
  {10.1146/annurev.astro.44.051905.092532}, \href
  {http://adsabs.harvard.edu/abs/2006ARA%26A..44...49R} {44, 49}

\bibitem[\protect\citeauthoryear{{Remillard}, {Morgan}, {McClintock}, {Bailyn}
  \& {Orosz}}{{Remillard} et~al.}{1999}]{Remillard99}
{Remillard} R.~A.,  {Morgan} E.~H.,  {McClintock} J.~E.,  {Bailyn} C.~D.,
  {Orosz} J.~A.,  1999, \mn@doi [\apj] {10.1086/307606}, \href
  {http://adsabs.harvard.edu/abs/1999ApJ...522..397R} {522, 397}

\bibitem[\protect\citeauthoryear{{Remillard}, {Muno}, {McClintock}  \&
  {Orosz}}{{Remillard} et~al.}{2002}]{Remillard02}
{Remillard} R.~A.,  {Muno} M.~P.,  {McClintock} J.~E.,   {Orosz} J.~A.,  2002,
  \mn@doi [\apj] {10.1086/343791}, \href
  {http://adsabs.harvard.edu/abs/2002ApJ...580.1030R} {580, 1030}

\bibitem[\protect\citeauthoryear{{Reynolds} \& {Miller}}{{Reynolds} \&
  {Miller}}{2009}]{Reynolds09}
{Reynolds} C.~S.,  {Miller} M.~C.,  2009, \mn@doi [\apj]
  {10.1088/0004-637X/692/1/869}, \href
  {http://adsabs.harvard.edu/abs/2009ApJ...692..869R} {692, 869}

\bibitem[\protect\citeauthoryear{{Shakura} \& {Sunyaev}}{{Shakura} \&
  {Sunyaev}}{1973}]{Shakura73}
{Shakura} N.~I.,  {Sunyaev} R.~A.,  1973, \aap, \href
  {http://adsabs.harvard.edu/abs/1973A%26A....24..337S} {24, 337}

\bibitem[\protect\citeauthoryear{{Silbergleit}, {Wagoner}  \&
  {Ortega-Rodr{\'{\i}}guez}}{{Silbergleit} et~al.}{2001}]{Silbergleit01}
{Silbergleit} A.~S.,  {Wagoner} R.~V.,   {Ortega-Rodr{\'{\i}}guez} M.,  2001,
  \mn@doi [\apj] {10.1086/318659}, \href
  {https://ui.adsabs.harvard.edu/abs/2001ApJ...548..335S} {548, 335}

\bibitem[\protect\citeauthoryear{{Strohmayer}}{{Strohmayer}}{2001}]{Strohmayer01}
{Strohmayer} T.~E.,  2001, \mn@doi [\apjl] {10.1086/320258}, \href
  {http://adsabs.harvard.edu/abs/2001ApJ...552L..49S} {552, L49}

\bibitem[\protect\citeauthoryear{{Wagoner}}{{Wagoner}}{1999}]{Wagoner99}
{Wagoner} R.~V.,  1999, \mn@doi [\physrep] {10.1016/S0370-1573(98)00104-5},
  \href {http://adsabs.harvard.edu/abs/1999PhR...311..259W} {311, 259}

\bibitem[\protect\citeauthoryear{{Wagoner}, {Silbergleit}  \&
  {Ortega-Rodr{\'{\i}}guez}}{{Wagoner} et~al.}{2001}]{Wagoner01}
{Wagoner} R.~V.,  {Silbergleit} A.~S.,   {Ortega-Rodr{\'{\i}}guez} M.,  2001,
  \mn@doi [\apjl] {10.1086/323655}, \href
  {https://ui.adsabs.harvard.edu/abs/2001ApJ...559L..25W} {559, L25}

\bibitem[\protect\citeauthoryear{{van der Klis}}{{van der
  Klis}}{2006}]{vanderKlis04}
{van der Klis} M.,  2006, {Rapid X-ray variability, Compact stellar X-ray
  sources, edited by Lewin, W.H.G. and van der Klis, M., pages 39-98,
  Cambridge, UK: Cambridge University Press}

\makeatother
\end{thebibliography}


\begin{thebibliography}{}
\expandafter\ifx\csname natexlab\endcsname\relax\def\natexlab#1{#1}\fi
\providecommand{\url}[1]{\href{#1}{#1}}

\bibitem[{{Abramowicz} \& {Klu{\'z}niak}(2001)}]{Abramowicz01}
{Abramowicz}, M.~A., \& {Klu{\'z}niak}, W. 2001, \aap, 374, L19

\bibitem[{{Bursa} {et~al.}(2004){Bursa}, {Abramowicz}, {Karas}, \&
  {Klu{\'z}niak}}]{Bursa04}
{Bursa}, M., {Abramowicz}, M.~A., {Karas}, V., \& {Klu{\'z}niak}, W. 2004,
  \apjl, 617, L45

\bibitem[{{Chen} \& {Taam}(1995)}]{Chen95}
{Chen}, X., \& {Taam}, R.~E. 1995, \apj, 441, 354

\bibitem[{{Fragile} {et~al.}(2018){Fragile}, {Etheridge}, {Anninos}, {Mishra},
  \& {Klu{\'z}niak}}]{Fragile18}
{Fragile}, P.~C., {Etheridge}, S.~M., {Anninos}, P., {Mishra}, B., \&
  {Klu{\'z}niak}, W. 2018, \apj, 857, 1

\bibitem[{{Giussani} {et~al.}(2014){Giussani}, {Klu{\'z}niak}, \&
  {Mishra}}]{Giussani15}
{Giussani}, L., {Klu{\'z}niak}, W., \& {Mishra}, B. 2014, in Proceedings of
  RAGtime 14-16: Workshop on blackholes and neutron stars, arXiv:1503.05546,
  ed. Z.~{Ztuchl{\'i}k}, G.~{T{\"o}r{\"o}k}, \& T.~{Pech{\'a}\v{c}ek}, 93--98

\bibitem[{{Homan} {et~al.}(2005){Homan}, {Miller}, {Wijnands}, {van der Klis},
  {Belloni}, {Steeghs}, \& {Lewin}}]{Homan05}
{Homan}, J., {Miller}, J.~M., {Wijnands}, R., {et~al.} 2005, \apj, 623, 383

\bibitem[{{Honma} {et~al.}(1992){Honma}, {Matsumoto}, \& {Kato}}]{Honma92}
{Honma}, F., {Matsumoto}, R., \& {Kato}, S. 1992, \pasj, 44, 529

\bibitem[{{Kato}(2001)}]{Kato2001}
{Kato}, S. 2001, \pasj, 53, 1

\bibitem[{{Kato}(2016)}]{Kato16}
---. 2016, {Oscillations of Disks}, Vol. 437, doi:10.1007/978-4-431-56208-5

\bibitem[{{Kato} {et~al.}(2008){Kato}, {Fukue}, \& {Mineshige}}]{Kato08}
{Kato}, S., {Fukue}, J., \& {Mineshige}, S. 2008, {Black-Hole Accretion Disks
  --- Towards a New Paradigm ---}

\bibitem[{{Klu{\'z}niak}(2005)}]{Kluzniak05}
{Klu{\'z}niak}, W. 2005, Astronomische Nachrichten, 326, 820

\bibitem[{{Kluzniak} \& {Abramowicz}(2001)}]{Kluzniak01}
{Kluzniak}, W., \& {Abramowicz}, M.~A. 2001, Acta Physica Polonica B, 32, 3605

\bibitem[{{Kluzniak} \& {Abramowicz}(2002)}]{Kluzniak02}
---. 2002, ArXiv Astrophysics e-prints, astro-ph/0203314

\bibitem[{{Klu{\'z}niak} {et~al.}(2004){Klu{\'z}niak}, {Abramowicz}, \&
  {Lee}}]{Kluzniak04}
{Klu{\'z}niak}, W., {Abramowicz}, M.~A., \& {Lee}, W.~H. 2004, in American
  Institute of Physics Conference Series, Vol. 714, X-ray Timing 2003: Rossi
  and Beyond, ed. P.~{Kaaret}, F.~K. {Lamb}, \& J.~H. {Swank}, 379--382

\bibitem[{{Klu{\'z}niak} \& {Lee}(2002)}]{KL2002}
{Klu{\'z}niak}, W., \& {Lee}, W.~H. 2002, \mnras, 335, L29

\bibitem[{{Mihalas} \& {Mihalas}(1984)}]{Mihalas84}
{Mihalas}, D., \& {Mihalas}, B.~W. 1984, {Foundations of radiation
  hydrodynamics}

\bibitem[{{Milsom} \& {Taam}(1996)}]{Milsom96}
{Milsom}, J.~A., \& {Taam}, R.~E. 1996, \mnras, 283, 919

\bibitem[{{Milsom} \& {Taam}(1997)}]{Milsom97}
---. 1997, \mnras, 286, 358

\bibitem[{{Mishra} {et~al.}(2018){Mishra}, {Klu{\'z}niak}, \&
  {Fragile}}]{Mishra18}
{Mishra}, B., {Klu{\'z}niak}, W., \& {Fragile}, P.~C. 2018, \mnras,
  arXiv:1810.05755

\bibitem[{{Mishra} {et~al.}(2017){Mishra}, {Vincent}, {Manousakis}, {Fragile},
  {Paumard}, \& {Klu{\'z}niak}}]{Mishra17}
{Mishra}, B., {Vincent}, F.~H., {Manousakis}, A., {et~al.} 2017, \mnras, 467,
  4036

\bibitem[{{Morgan} {et~al.}(1997){Morgan}, {Remillard}, \&
  {Greiner}}]{Morgan97}
{Morgan}, E.~H., {Remillard}, R.~A., \& {Greiner}, J. 1997, \apj, 482, 993

\bibitem[{{Nowak} \& {Wagoner}(1991)}]{Nowak91}
{Nowak}, M.~A., \& {Wagoner}, R.~V. 1991, \apj, 378, 656

\bibitem[{{Nowak} \& {Wagoner}(1992)}]{Nowak92}
---. 1992, \apj, 393, 697

\bibitem[{{Nowak} \& {Wagoner}(1993)}]{Nowak93}
---. 1993, \apj, 418, 187

\bibitem[{{Okazaki} {et~al.}(1987){Okazaki}, {Kato}, \& {Fukue}}]{Okazaki87}
{Okazaki}, A.~T., {Kato}, S., \& {Fukue}, J. 1987, \pasj, 39, 457

\bibitem[{{O'Neill} {et~al.}(2009){O'Neill}, {Reynolds}, \&
  {Miller}}]{ONeill09}
{O'Neill}, S.~M., {Reynolds}, C.~S., \& {Miller}, M.~C. 2009, \apj, 693, 1100

\bibitem[{{Perez} {et~al.}(1997){Perez}, {Silbergleit}, {Wagoner}, \&
  {Lehr}}]{Perez97}
{Perez}, C.~A., {Silbergleit}, A.~S., {Wagoner}, R.~V., \& {Lehr}, D.~E. 1997,
  \apj, 476, 589

\bibitem[{{Remillard}(2004)}]{Remillard04}
{Remillard}, R.~A. 2004, in American Institute of Physics Conference Series,
  Vol. 714, X-ray Timing 2003: Rossi and Beyond, ed. P.~{Kaaret}, F.~K. {Lamb},
  \& J.~H. {Swank}, 13--20

\bibitem[{{Remillard} {et~al.}(1999){Remillard}, {Morgan}, {McClintock},
  {Bailyn}, \& {Orosz}}]{Remillard99}
{Remillard}, R.~A., {Morgan}, E.~H., {McClintock}, J.~E., {Bailyn}, C.~D., \&
  {Orosz}, J.~A. 1999, \apj, 522, 397

\bibitem[{{Remillard} {et~al.}(2002){Remillard}, {Muno}, {McClintock}, \&
  {Orosz}}]{Remillard02}
{Remillard}, R.~A., {Muno}, M.~P., {McClintock}, J.~E., \& {Orosz}, J.~A. 2002,
  \apj, 580, 1030

\bibitem[{{Reynolds} \& {Miller}(2009)}]{Reynolds09}
{Reynolds}, C.~S., \& {Miller}, M.~C. 2009, \apj, 692, 869

\bibitem[{{Strohmayer}(2001)}]{Strohmayer01}
{Strohmayer}, T.~E. 2001, \apjl, 552, L49

\bibitem[{{van der Klis}(2006)}]{vanderKlis04}
{van der Klis}, M. 2006, {Rapid X-ray variability, Compact stellar X-ray
  sources, edited by Lewin, W.H.G. and van der Klis, M., pages 39-98,
  Cambridge, UK: Cambridge University Press}

\bibitem[{{Wagoner}(1999)}]{Wagoner99}
{Wagoner}, R.~V. 1999, \physrep, 311, 259

\end{thebibliography}

\end{document}